\pdfoutput=1
\RequirePackage[T1]{fontenc}

\documentclass[11pt]{article} 
\usepackage[height=8.85in,width=6.45in]{geometry}

\usepackage{amsmath}
\usepackage{mathrsfs}
\usepackage{amssymb}
\usepackage{mathtools}
\usepackage{slashed}
\usepackage{physics,braket}
\usepackage{graphicx,rotating}
\usepackage{xcolor}
\usepackage[utf8]{inputenc}
\usepackage{acronym}
\usepackage{bm}
\usepackage{bbm}
\usepackage{cite}
\usepackage{times}
\usepackage{mdframed}
\usepackage{datetime}
\usepackage{dsfont}
\usepackage{multirow}
\usepackage{cancel}
\usepackage{caption}
\usepackage{subcaption}
\usepackage{pifont}
\usepackage{romannum}
\usepackage{chngcntr}
\usepackage[most]{tcolorbox}
\usepackage{leftindex}

\definecolor{darkred}{rgb}{0.7, 0., 0.}
\definecolor{mblue}{HTML}{0019A2}
\definecolor{monza}{HTML}{CF000F}
\definecolor{darkmagenta}{HTML}{8b008b}
\usepackage[bookmarksopen,
colorlinks=true,
linkcolor=darkred,
menucolor=mblue,
anchorcolor=mblue,
citecolor=mblue,
urlcolor=mblue,
linktoc=all]{hyperref}

\AtBeginDocument{\pagenumbering{arabic}}
\numberwithin{equation}{section}

\renewcommand{\arraystretch}{1.5}

\def\ini{{\mathrm{ini}}}

\def\hGamma{\hat{\Gamma}}

\def\hhT{\hat{h}^{\mathrm{T}}}

\def\ch{\widetilde{h}}

\def\cV{\widetilde{V}}
\def\VT{V^{\mathrm{T}}}
\def\cVT{\cV^{\mathrm{T}}}

\def\cS{\widetilde{S}} 
\def\ST{S^{\mathrm{T}}}
\def\cST{\widetilde{S}^{\mathrm{T}}}

\def\Hz{{~\mathrm{Hz}}}
\def\hmode{{h_{(\lambda)}}}

\newcommand{\ee}{\mathrm{e}}

\newcommand{\cv}{c_{\mathrm{v}}}

\newcommand{\calH}{\mathcal{H}}

\newcommand{\calP}{\mathcal{P}}

\newcommand{\eg}{\textit{e.g.}}

\newcommand{\muk}{\mu_k}

\newcommand{\DDD}{\delta^{(3)}_D}
\newcommand{\DD}{\delta^{(2)}_D}
\newcommand{\Pt}{P^{(\mathrm{T})}}
\newcommand{\Pv}{P^{(\mathrm{V})}}

\DeclareMathOperator{\diag}{diag}

\newcommand{\bae}[1]{\begin{align} #1 \end{align}}

\newcommand{\beae}[1]{\begin{equation}\begin{aligned} #1 \end{aligned}\end{equation}}
\newcommand{\baea}[1]{\begin{align} \begin{aligned} #1 \end{aligned}\end{align}}

\newcommand{\vece}{e}
\newcommand{\hatk}{\hat{k}}

\newcommand{\Hquad}{\hspace{.25em}} 

\usepackage{relsize}
\newcommand{\Hugelike}{\fontsize{24pt}{30pt}\selectfont}

\acrodef{EoM}{equation of motion}
\acrodef{CMB}{cosmic microwave background}
\acrodef{ORF}{overlap reduction function}
\acrodef{GWs}{gravitational waves}
\acrodef{GW}{gravitational wave}
\acrodef{LOS}{line of sight}
\acrodef{EFT}{effective field theory}
\acrodef{PTA}{pulsar timing arrays}
\acrodef{HD}{Hellings--Downs}
\acrodef{GR}{general relativity}
\acrodef{IAs}{intrinsic alignments}

\begin{document}

\begin{titlepage}

\begin{center}	
  \hfill 
\makebox[0pt][r]{YITP-25-116}
\vskip .6in
{\Hugelike \bfseries
      Imprints of gravitational-wave polarizations 
 \\ \vskip .2in
       on projected tidal tensor in three dimensions
    }	
    \vskip .8in
    {\LARGE
        Yusuke Mikura,$^{\ast,}$\footnote{\href{mailto:ymikura@asiaa.sinica.edu.tw}{\color{darkred}{ymikura@asiaa.sinica.edu.tw}}}
        \Hquad
        Teppei Okumura,$^{\ast \dag,}$\footnote{\href{mailto:tokumura@asiaa.sinica.edu.tw}{\color{darkred}{tokumura@asiaa.sinica.edu.tw}}}
        \Hquad
        Misao Sasaki$^{\dag \ddag \S \$,}$\footnote{\href{mailto:misao.sasaki@ipmu.jp}{\color{darkred}{misao.sasaki@ipmu.jp}}}
    }
    \vskip 1.cm

    \def\arraystretch{1}
	\begin{tabular}{ll}
      $^\ast$& 
      Institute of Astronomy and Astrophysics, Academia Sinica, \\
      & No. 1, Section 4, Roosevelt Road, Taipei 10617, Taiwan
      \\[0.5em]
      $^\dag$& 
      Kavli Institute for the Physics and Mathematics of the Universe (WPI), UTIAS, \\
      & The University of Tokyo, Kashiwa, Chiba 277-8583, Japan
      \\[0.5em]
      $^\ddag$& 
      Asia Pacific Center for Theoretical Physics (APCTP),
      \\
      & Pohang 37673, Korea
      \\[0.5em]
      $^\S$& 
      Center for Gravitational Physics, Yukawa Institute for Theoretical Physics, 
      \\
      & Kyoto University, Kyoto 606-8502, Japan
      \\[0.5em]
      $^\$ $& 
      Leung Center for Cosmology and Particle Astrophysics,
      \\
      & National Taiwan University, Taipei 10617, Taiwan

	\end{tabular}
	\vskip 1.cm
\end{center}

\noindent\ignorespaces
{\bfseries Abstract:}
Gravitational waves (GWs) distort galaxy shapes through the tidal effect, offering a novel avenue to probe the nature of gravity. In this paper, we investigate how extra GW polarizations beyond those predicted by general relativity imprint observable signatures on galaxy shapes. 
Since galaxy shapes are measured as two-dimensional images projected onto the celestial sphere, we present three-dimensional statistical quantities of the projected tidal tensor sourced by the tensor perturbation.
We show that the presence of extra polarization modes modifies both the amplitude and angular dependence of the correlation functions. Furthermore, we identify a distinct observational channel for probing parity violation in helicity-two and helicity-one modes. In particular, we show that if they propagate at different speeds, galaxy surveys can disentangle the source of parity violation. Our findings establish a theoretical framework for using upcoming large-scale galaxy surveys to test modified gravity theories through the polarization content of GWs.
\end{titlepage}

\setcounter{tocdepth}{3}
{
  \hypersetup{linkcolor=darkred}
  \tableofcontents
}

\section{Introduction}
The detection of \ac{GWs} by the LIGO-Virgo Collaboration has opened a new era of cosmology where we can probe fundamental physics through \ac{GWs}~\cite{LIGOScientific:2016aoc,LIGOScientific:2016lio,LIGOScientific:2016sjg,LIGOScientific:2016dsl,Yunes:2016jcc}. There are many mechanisms that can generate a stochastic GW background over a huge range of frequencies, and appropriate ways to search for the signals vary depending on the frequency of interest. The lowest frequency band accessible to us spans approximately from $10^{-20}\Hz$ to $10^{-16}\Hz$, where \ac{GWs} originate in the vacuum fluctuations during inflation.
This frequency range can be probed by the $B$-mode signal in the polarization of the cosmic microwave background, and it has been a central topic in observational cosmology for decades. Other than that, there exist several GW observatories searching for signals with frequencies ranging from $10^{-9}\Hz$ to $10^{4}\Hz$. For instance, ground-based observatories probe the high-frequency \ac{GWs} ranging from $10\Hz$-$10^{4}\Hz$~\cite{KAGRA:2013rdx,VIRGO:2014yos,LIGOScientific:2014pky,TianQin:2015yph} and space-based detectors will observe \ac{GWs} with frequencies between $10^{-4}\Hz$ and $10^{-1}\Hz$~\cite{LISA:2017pwj,Kawamura:2020pcg}. It is also beneficial to utilize \ac{PTA} to explore the spectrum between $10^{-9}\Hz$ and $10^{-6}\Hz$, where of particular interest would go to the recent evidence reported by the NANOGrav collaboration~\cite{NANOGrav:2020bcs}. Even with the above-mentioned observatories, there still exists a wide range of window that has not been explored well. Interestingly, some of the frequency windows, roughly around $10^{-16}\Hz$ to $10^{-14}\Hz$, correspond to the scale of the large-scale structure of the universe, thereby having a potential to be investigated observationally in light of the wealth of upcoming surveys. 

There are a variety of ways to extract information about fundamental physics from the large-scale structure. Among several observables, galaxy-shape correlations, referred to as \ac{IAs}~\cite{Lee:1999ii,Catelan:2000vm,Crittenden:2000wi,Hirata:2004gc}, can be beneficial because galaxy shapes are intrinsically aligned with the surrounding environment. Given that the universe is homogeneous and isotropic on large scales, orientations of galaxies were expected to be random, so that the \ac{IAs} were first considered as one of the major sources of systematic errors in weak-lensing measurements (see, \eg, refs.~\cite{Troxel:2014dba,Joachimi:2015mma} and references therein). However, it has been recently realized that the \ac{IAs} themselves have rich information to constrain cosmological parameters, with the large-scale dark matter distribution being a primary example. On large scales, we often adopt the so-called linear alignment model where the intrinsic galaxy shape is assumed to respond linearly to the tidal field generated by the gravitational potential~\cite{Catelan:2000vm,Hirata:2004gc,Blazek:2011xq,Okumura:2019ozd}. This model has been confirmed in both simulations and observations~\cite{Okumura:2008bm,Okumura:2008du,Blazek:2011xq}, implying that the \ac{IAs} can be indeed an interesting and practical tool to extract cosmological information~\cite{Chisari:2013dda,Schmidt:2015xka,Chisari:2016xki,Kogai:2020vzz,Lee:2022lbu}. 
The amount of cosmological information can be maximized by combining recent theoretical developments~\cite{Okumura:2019ozd,Okumura:2019ned,Okumura:2020hhr,Akitsu:2020jvx,Akitsu:2020fpg,Shiraishi:2020vvj,Shi:2021pqp,Kurita:2022agh} with measurements of galaxy shapes via imaging observations and of distances to galaxies via spectroscopic observations\cite{Okumura:2019ozd,Taruya:2020tdi,Kurita:2020hap,Okumura:2021xgc,Chuang:2021evv,Saga:2022frj,Shiraishi:2023zda,Okumura:2023pxv,Kurita:2023qku,Matsubara:2023avg,Okumura:2025xnv}.

The origin of the intrinsic shape distortion is not limited to the gravitational potential; it can also be sourced by a rank-two tensor field at leading order~\cite{Schmidt:2013gwa}.\footnote{The effects of the tensor field on the galaxy clustering and weak lensing have been studied in refs.~\cite{Kaiser:1996wk,Dodelson:2003bv,Cooray:2005hm,Yoo:2009au,Dodelson:2010qu,Masui:2010cz,Schmidt:2012ne,Jeong:2012nu,Schmidt:2012nw,Chisari:2014xia}.}
This comes from the fact that the tidal force is associated with the perturbed Riemann tensor that is linearly proportional to the tensor perturbation. This idea was generalized to include the effects of the chirality of tensor modes of \ac{GWs} on angular statistics in ref.~\cite{Biagetti:2020lpx}. 
To compare with observations, one has to deal with the linear shape bias for \ac{GWs} as in the case of the galaxy bias. Ref.~\cite{Akitsu:2022lkl} tested the ansatz of the linear shape bias used in ref.~\cite{Schmidt:2013gwa} and confirmed its validity using $N$-body simulations.
Subsequently, this line of research was further extended by examining the auto- and cross-power spectra of the shear $E$- and $B$-modes induced by the scalar, vector, and tensor modes in the three-dimensional space~\cite{Philcox:2023uor,Saga:2023afb}. Two of the authors provided in ref.~\cite{Okumura:2024xnd} a complementary technique to extract signals of the tensor modes using real-space observables, where a primary quantity is an analogue of the \ac{ORF} which describes the angular dependence of signals from sources that are spatially separated on the celestial sphere. Ref.~\cite{Okumura:2024xnd} showed that, due to the propagation,  signals of the tensor modes in the \ac{ORF} have an oscillatory feature which is absent in ones of the scalar tidal force, making it possible to extract information about the tensor modes out of signals that the scalar tidal force dominates.

In this paper we generalize ref.~\cite{Okumura:2024xnd} by including additional scalar and vector components that may appear in metric theories of gravity beyond \ac{GR}~\cite{Eardley:1973br,Eardley:1973zuo}. Classic examples of such include scalar-tensor theories, massive gravity theories, Einstein--{\AE}ther theories, and scalar-tensor-vector gravity. 
Considering upcoming galaxy surveys, it is informative to understand how additional polarization modes affect the analogue of the \ac{ORF} because it serves as the foundations for fitting some modified-gravity theories with observational data.\footnote{In \ac{PTA}, the \ac{ORF} in \ac{GR} is known as the \ac{HD} curve~\cite{Hellings:1983fr} and, since deviations from the \ac{HD} curve is a hint of new physics beyond \ac{GR}, a number of investigations have been made observationally~\cite{NANOGrav:2023ygs} and theoretically~\cite{Liang:2023ary,Liang:2024mex,Cordes:2024oem,Hu:2024wub}.}

The paper is organized as follows. In section~\ref{sec. Correlation functions of gravitational waves}, we define the polarization tensors and the power spectra of \ac{GWs} to fix the overall notation. 
In section~\ref{Projection onto the celestial sphere}, we turn to a technique to extract signals of \ac{GWs} using the projection of a tensor field onto the celestial sphere, presenting general calculations for the power spectra of the traceless projected tensor field. 
We then investigate the angular- and distance-dependence of an analogue of the \ac{ORF} with three concrete examples in section~\ref{sec. ORF}.
Finally, we conclude in section~\ref{sec. Conclusion}. 
In appendix~\ref{sec. Power spectra of the tensor perturbation}, we provide a list of correlations for the projected tensor field without extracting trace. In appendix~\ref{sec. EB correlation}, we discuss a relation between our formalism and the $E/B$-mode decomposition. In appendix~\ref{sec. the angular integral}, we provide a different way of calculating the analogue of the \ac{ORF}.
Throughout the paper, we adopt the natural unit $c=1$ and use $\eta_{\mu\nu} = \diag(-1,1,1,1)$ for the sign of the Minkowski metric. A bar denotes a complex conjugate. We use the lowercase Latin alphabet for spacial components in three dimensions and the uppercase Latin alphabet for components in the two-dimensional projected space.

\section{Power spectra of gravitational waves}\label{sec. Correlation functions of gravitational waves}
We define \ac{GWs} as the spatial perturbation obtained after imposing the synchronous gauge $h_{0\mu}=0$ on the tensor perturbation $h_{\mu\nu}$. The line element is then given by
\bae{\label{eq. metric}
  \dd s^2 = a^{2}(\eta)\left[-\dd \eta^2 + \left(\delta_{ij}+h_{ij}\right)\dd x^i \dd x^j\right] ~,
} 
where $a$ is the scale factor and $\eta$ is the conformal time. 
In three-dimensional space, the tensor perturbation $h_{ij}$ can accommodate up to six helicity modes~\cite{Eardley:1973br,Eardley:1973zuo}, that can be labeled by an index $\lambda=\{\pm 2,\pm 1, b, \ell\}$ in the chiral basis. The last two modes have helicity-zero and are called breathing and longitudinal polarizations, respectively. In the Fourier space, the tensor perturbation can be written as
\bae{\label{Eq. Polarization tensors}
h_{ij} (\eta, \bm{k}) = \sum_{\lambda} \vece^{(\lambda)}{}_{ij} \hmode (\eta, \bm{k})
~,
}
where the polarization tensors $\vece^{(\lambda)}{}_{ij}$ are defined as~\cite{Takeda:2018uai}
\bae{
\vece^{(\pm 2)}{}_{ij} & \coloneqq \frac12 \left[\left(\vece^{(1)}{}_i \otimes \vece^{(1)}{}_j -\vece^{(2)}{}_i\otimes \vece^{(2)}{}_j\right) \mp i \left(\vece^{(1)}{}_i \otimes \vece^{(2)}{}_j + \vece^{(2)}{}_i\otimes \vece^{(1)}{}_j \right)\right] ~,
\\
\vece^{(\pm 1)}{}_{ij} & \coloneqq \frac12 \left[\left(\vece^{(1)}{}_i \otimes \hatk_j +\hatk_i\otimes \vece^{(1)}{}_j\right) \mp i \left(\vece^{(2)}{}_i \otimes \hatk_j + \hatk_i\otimes \vece^{(2)}{}_j \right)\right] ~,
\\
\vece^{(b)}{}_{ij} & \coloneqq \frac{1}{\sqrt{2}}\left(\vece^{(1)}{}_i \otimes \vece^{(1)}{}_j + \vece^{(2)}{}_i\otimes \vece^{(2)}{}_j \right) ~,
\\
\vece^{(\ell)}{}_{ij} & \coloneqq \hatk_i \otimes \hatk_j ~.
}
Here, $\hat{\bm{k}}$ denotes a unit vector pointing to the propagation of the tensor perturbation, and $\bm{e}^{(1)}$ and $\bm{e}^{(2)}$ are two orthonormal vectors. The set of unit vectors $\{\bm{e}^{(1)}, \bm{e}^{(2)}, \hat{\bm{k}}\}$ forms the right-handed Cartesian coordinate system. One can easily show that the above polarizations satisfy
\bae{
\vece^{(\lambda)}{}_{ij}\bar{\vece}^{(\lambda^\prime)}{}^{ij} = \delta^{\lambda \lambda^\prime} ~,
}
where the bar denotes the complex conjugate and the polarization tensors with upper indices are the same as those with lower indices, namely $\vece^{(\lambda)}{}^{ij}=\vece^{(\lambda)}{}_{ij}$. We use the Einstein summation convention throughout the paper. 

The temporal power spectrum of \ac{GWs}, $P_h (\eta, \eta^\prime, k)$, is defined by the ensemble average as
\bae{
\Braket{h^{ij} (\eta, \bm{k}) h_{ij} (\eta^\prime, \bm{k}^\prime)} = \left(2\pi\right)^3\delta^{(3)}_D (\bm{k}+\bm{k}^\prime) P_h (\eta, \eta^\prime, k) ~,
}
where $\delta^{(n)}_D$ is the $n$-dimensional Dirac delta function. Under the assumption of an isotropic stochastic GW background, one expects that the cross-power spectra between different helicity modes vanish, thereby giving the following power spectrum for each polarization:
\bae{
\Braket{h_{(\lambda)} (\eta, \bm{k}) h_{(\lambda^\prime)} (\eta^\prime, \bm{k}^\prime)} = \left(2\pi\right)^3\DDD (\bm{k}+\bm{k}^\prime) \delta_{\lambda\lambda^\prime}P^{(\lambda)} (\eta, \eta^\prime, k) ~.
}
For later convenience, let us define tensor- and vector-type power spectra as
\bae{
\Pt \coloneqq P^{(+2)} + P^{(-2)} ~, \quad \Pv \coloneqq P^{(+1)} + P^{(-1)} ~,
}
with which the total power spectrum $P_h (\eta, \eta^\prime, k)$ can be expressed as
\bae{
 P_h (\eta, \eta^\prime, k) = \Pt (\eta, \eta^\prime, k)+ \Pv (\eta, \eta^\prime, k) + P^{(b)} (\eta, \eta^\prime, k) + P^{(\ell)} (\eta, \eta^\prime, k) ~.
}
For the helicity-two or helicity-one modes, there can be an asymmetry of amplitudes between left- and right-handed waves. To parametrize this chirality, we define tensor- and vector-type chiral parameters $\chi^{(\mathrm{T})}$ and $\chi^{(\mathrm{V})}$ as
\bae{ \label{eq. chirality parameters}
\chi^{(\mathrm{T})} \coloneqq \frac{P^{(+2)}-P^{(-2)}}{\Pt} ~, 
\quad
\chi^{(\mathrm{V})} \coloneqq \frac{P^{(+1)}-P^{(-1)}}{\Pv} ~.
}

We next summarize the sub-horizon dynamics of the tensor perturbation. The wave equation for the mode function $h_{(\lambda)}(\eta, \bm{k})$ is given by
\bae{
 h_{(\lambda)}^{\prime\prime} +2\calH h_{(\lambda)}^{\prime} +c_{\lambda}^2 k^2 h_{(\lambda)} = 0 ~,
}
where primes denote derivatives with respect to the conformal time $\eta$, $\calH= a^{\prime}/a$ is the conformal Hubble parameter, and $c_{\lambda}$ is the propagation speed of the mode with helicity $\lambda$. 
Here and hereafter, we assume that the propagation
speed is nearly constant but can be different from the speed of light. Deep inside the horizon, a solution to the wave equation is given by
\bae{
\hmode(\eta, \bm{k}) \simeq \frac{a(\eta_\ini)}{a(\eta)} \left[\cos\left[c_{\lambda} k (\eta-\eta_\ini) + \phi_{\bm{k}} \right] h_{(\lambda)} (\eta_\ini, \bm{k}) + \frac{1}{c_{\lambda} k}\sin\left[c_{\lambda} k (\eta-\eta_\ini) + \phi_{\bm{k}} \right] h_{(\lambda)}^{\prime} (\eta_\ini, \bm{k}) \right] ~,
}
where $\eta_\ini$ denotes the conformal time either right after \ac{GWs} are generated or when the scale entered the horizon if GWs are primordial.
The phase $\phi_{\bm{k}}$ represents the effect of inhomogeneities during propagation from $\eta_\ini$ to $\eta$. We note that our discussion is applicable to both \ac{GWs} with primordial origin and ones generated by some other sources.
By generation we mean that the  \ac{GWs} are produced by causal sub-horizon mechanisms including astrophysical events and cosmological events such as phase transitions.
If \ac{GWs} are of the primordial origin, they may be phase coherent at first. 
However, long-distance propagation on the perturbed universe randomizes the phase $\phi_k$, allowing us to assume it random~\cite{Allen:1999xw,Margalit:2020sxp}. Therefore, in both cases, the power spectrum can be written as
\bae{
P^{(\lambda)} (\eta, \eta^\prime, k) 
\simeq \frac{a^2 (\eta_\ini)}{a(\eta) a(\eta^\prime)}
\cos\left[c_{\lambda} k \left(\eta-\eta^\prime\right)\right] P^{(\lambda)}_\ini (k) ~, \label{eq. Power dynamics}
}
where $P^{(\lambda)}_\ini (k)$ is the initial power spectrum.

We finally note that the galaxy-shape distortion is induced not by the tensor perturbation itself but by its time derivatives~\cite{Schmidt:2012nw}, namely, 
\bae{
t_{ij} \propto -\frac{2}{3\calH^2}\delta R_{i0j0} = - \frac{1}{3\calH^2}\left(h_{ij}^{\prime\prime}+\calH h_{ij}^{\prime}\right) ~.
}
Inside the horizon, this can be simplified by the use of the wave equation as
\bae{\label{eq. shape field and tensor}
  t_{ij} \propto -\frac{k^2}{3\calH^2} \hat{h}_{ij} (\eta, \bm{k}) ~,
}
where we introduce a new field $\hat{h}_{ij}$ by
\bae{\label{eq. tensor observable}
 \hat{h}_{ij} (\eta, \bm{k}) \coloneqq \sum_{\lambda} c_{\lambda}^2 \vece^{(\lambda)}{}_{ij} \hmode (\eta, \bm{k}) ~.
}
This field is different from the original tensor perturbation when propagating speed of modes is not identical to the speed of light. In the following, we call $\hat{h}_{ij}$ the tensor perturbation.
\section{Power spectra of projected traceless tensor}\label{Projection onto the celestial sphere}
Given the fact that galaxy shapes are observed as two-dimensional images from galaxy surveys, we need to introduce correlation functions in three dimensions using the tensor perturbation projected onto the celestial sphere. In this work, we focus on its trace-free part with an expectation that its trace part cannot be easily evaluated in the presence of an extrinsic (lensing) contribution. In section~\ref{sec. traceless Power spectra on the projected space}, we explain how the tensor perturbation is projected onto the celestial sphere. We then present the power spectra made up with tensors with trace extracted in subsequent sections.\footnote{See appendix~\ref{sec. Power spectra of the tensor perturbation} for the power spectra of the tensor perturbation without extracting trace and appendix~\ref{sec. EB correlation} for the relation between our formalism and the $E/B$-mode decomposition.}

\subsection{Projection onto the two-dimensional plane}\label{sec. traceless Power spectra on the projected space}
Let us start with the projection of the tensor perturbation~\eqref{eq. tensor observable} onto a two-dimensional space under the flat-sky approximation. To this end, we write a coordinate system in real space as $\{\hat{\bm{x}}, \hat{\bm{y}}, \hat{\bm{z}}\}$, and choose such that the $z$-axis points to the \ac{LOS} and the $x$-axis is aligned with $\bm{e}^{(1)}$ which is the orthonormal vector of the system $\{\bm{e}^{(1)}, \bm{e}^{(2)}, \hat{\bm{k}}\}$. In the system $\{\hat{\bm{x}}, \hat{\bm{y}}, \hat{\bm{z}}\}$, the vectors $\bm{e}^{(1)}$, $\bm{e}^{(2)}$, and $\hat{\bm{k}}$ can be expressed as
\bae{\label{eq. orthonoraml wavevector}
\vece^{(1)}{}_i = \left( 1, 0, 0 \right) ~,
\quad
\vece^{(2)}{}_i = \left( 0, \cos{\theta_k}, -\sin{\theta_k} \right) ~,
\quad
\hat{k}_i = \left( 0, \sin{\theta_k}, \cos{\theta_k} \right) ~,
}
where $\theta_k$ is the angle between the propagating direction of \ac{GWs} and \ac{LOS}. The two-dimensional space, which we call the projected space, is spanned by the first and second components in the Cartesian coordinate system $\{\hat{\bm{x}}, \hat{\bm{y}}\}$. Therefore, components of the orthonormal vectors $\bm{e}^{(1)}$, $\bm{e}^{(2)}$, and $\hat{\bm{k}}$ in the projected space can be obtained by dropping the third component as
\bae{\label{eq. projection wavevector}
\vece^{(1)}{}_A = \left( 1, 0\right) ~,
\quad
\vece^{(2)}{}_A = \left( 0, \cos{\theta_k} \right) ~,
\quad
\hat{k}_A = \left( 0, \sin{\theta_k} \right) ~,
}
where we use the uppercase Latin alphabet for components in the projected space.
With use of the projected components~\eqref{eq. projection wavevector}, the polarization tensors can be written as
\bae{
  \vece^{(\lambda)}{}_{ij} = \mqty(\vece^{(\lambda)}{}_{AB} & \vece^{(\lambda)}{}_{A z} \\
  \vece^{(\lambda)}{}_{z B} & \vece^{(\lambda)}{}_{zz}) ~,
}
with which one can define the projected tensor field as
  \bae{
    \hat{h}_{AB} (\eta, \bm{k}) = \sum_{\lambda} c_{\lambda}^2 \vece^{(\lambda)}{}_{AB} \hmode (\eta, \bm{k}) ~.
}
One can introduce lower rank tensors out of $\hat{h}_{AB}$, and we call them projected vector or scalar fields depending on the number of indices. It should be emphasized that we are interested in correlation functions in three dimensions. 

The trace part of the projected tensor field can be easily removed as
\bae{\label{eq. traceless projected tensor}
\hat{h}^{\mathrm{T}}{}_{AB} \coloneqq \hat{h}_{AB}-\frac{1}{2}\delta_{AB} \hat{h}^{C}{}_C ~,
}
where $\delta_{AB}$ is the Kronecker delta in two dimensions. Let us take a look at how each component of the traceless tensor~\eqref{eq. traceless projected tensor} is related to the mode functions defined in the three-dimensional space. 
There exist two components in the traceless projected tensor field since it is a symmetric $2\times 2$ matrix
\bae{\label{eq. matrix of traceless projected tensor}
\hat{h}^{\mathrm{T}}{}_{AB} (\eta, \bm{k}) = \mqty(
  \hat{h}^{\mathrm{T}}{}_{xx} & \hat{h}^{\mathrm{T}}{}_{xy} \\
  \hat{h}^{\mathrm{T}}{}_{yx} & -\hat{h}^{\mathrm{T}}{}_{xx}
  ) ~.
}
The $(x,x)$ components is given by
\beae{
    \hat{h}^{\mathrm{T}}{}_{xx} = ~& \frac14 \bigg[ 
    \left(1+\cos^2{\theta_k}\right) \left(c_{+2}^2 h_{(+2)} + c_{-2}^2 h_{(-2)}\right)
    +i \sin{\left(2\theta_k\right)} \left(c_{+1}^2 h_{(+1)} - c_{-1}^2 h_{(-1)}\right)
    \\
    & +\sqrt{2} \sin^2{\theta_k} c_{b}^2 h_{(b)}
    - 2\sin^2{\theta_k} c_{\ell}^2 h_{(\ell)} \bigg] ~,
}
and the off-diagonal component is composed of the helicity-two and helicity-one modes as
\bae{
  \hat{h}^{\mathrm{T}}{}_{xy} = \hat{h}^{\mathrm{T}}{}_{yx} = \frac12 \left[ - i \cos{\theta_k}  \left(c_{+2}^2 h_{(+2)} - c_{-2}^2 h_{(-2)}\right) 
  + \sin{\theta_k} \left(c_{+1}^2 h_{(+1)} + c_{-1}^2 h_{(-1)}\right)\right] ~.
  }
It is obvious from above expressions that multiplication of the diagonal or off-diagonal components themselves give rise to parity-even power spectra. In other words, if parity violation exists, its signals appear only in power spectra that mix the diagonal and off-diagonal components. To see these, in the following, we explicitly provide a list of power spectra of fields that are constructed from the traceless projected tensor.
\subsection{Projected tensor field}
The auto-correlation of the traceless projected tensor field can be written as
\bae{
\Braket{\hhT{}^{AB}(\eta, \bm{k}) \hhT{}_{AB} (\eta^\prime, \bm{k}^\prime)} = \left(2\pi\right)^3\DD  (\bm{k}+\bm{k}^\prime) F_{\hat{h}\hat{h}}^{\mathrm{T}} (\muk, \eta, \eta^\prime, k) ~.
}
The function $F_{\hat{h}\hat{h}}^{\mathrm{T}} (\muk, \eta, \eta^\prime, k)$ depends on the angle between the propagating direction of \ac{GWs} and \ac{LOS}, $\muk\coloneqq\cos{\theta_k}$, and the power spectra defined in section~\ref{sec. Correlation functions of gravitational waves}. Its explicit form can be obtained straightforwardly as
\beae{\label{eq. F function tensor}
F_{\hat{h}\hat{h}}^{\mathrm{T}} (\muk, \eta, \eta^\prime, k) = ~& \frac18 \left(1+6\muk^2+\muk^4\right)
\left\{c_{+2}^2 P^{(+2)} (\eta, \eta^\prime, k) + c_{-2}^2 P^{(-2)} (\eta, \eta^\prime, k)\right\}
\\
&
+ \frac12 \left(1 - \muk^4\right)\left\{c_{+1}^2 P^{(+1)} (\eta, \eta^\prime, k) + c_{-1}^2 P^{(-1)} (\eta, \eta^\prime, k)\right\}
\\
& + \frac14 \left(1 - \muk^2\right)^2 c_{b}^2 P^{(b)}(\eta, \eta^\prime, k)
+ \frac12 \left(1 - \muk^2\right)^2 c_{\ell}^2 P^{(\ell)}(\eta, \eta^\prime, k)
~,
}
where the $\muk$-dependence for the helicity-$2$ mode coincides with eq.~(3.5) of ref.~\cite{Okumura:2024xnd}.
We can see from above that this correlation is parity-even, which is as expected since it is obtained as a combination of $(\hhT{}_{xx})^2$ and $(\hhT{}_{xy})^2$. We also observe that the breathing and longitudinal modes have the same $\muk$-dependence in their coefficients as their polarization tensors satisfy
\bae{
  \abs{\vece^{(b)}_{AB}-\frac12 \delta_{AB}\left(\vece^{(b)}\right)^C{}_{C}}^2 = \frac12 \abs{\vece^{(\ell)}_{AB}-\frac12 \delta_{AB}\left(\vece^{(\ell)}\right)^C{}_{C}}^2 ~.
}
This holds true as long as we consider projected quantities. However, the breathing and longitudinal modes can have different coefficients if the trace part is taken into account (see appendix~\ref{sec. Power spectra of the tensor perturbation}).
\subsection{Projected vector field}\label{sec. Projected vector perturbation}
In addition to the quadrupole moments corresponding to the shear, it may be possible to extract vector quantities from galaxy images~\cite{Goldberg:2004hh}. Expecting this, we consider the power spectra constructed with a derivative of the projected tensor field with trace extracted.

Out of the traceless projected tensor, we can define two vectors by\footnote{We do not use the hat as these quantities would not be confused with the tensor perturbation.}
\bae{\label{eq. vector without trace}
\VT{}_{A} (\eta, \bm{k}) & \coloneqq i k^{B} \hhT{}_{BA}(\eta, \bm{k}) = \left(i k\sin{\theta_k}\hhT{}_{yx}, - i k\sin{\theta_k}\hhT{}_{xx}\right) ~,
\\ \label{eq. curl vector without trace}
\cVT{}_A(\eta, \bm{k}) & \coloneqq i \epsilon^{BC}k_{B} \hhT{}_{CA}(\eta, \bm{k}) = \left(i k\sin{\theta_k}\hhT{}_{xx}, i k\sin{\theta_k}\hhT{}_{xy}\right) ~,
}
where $\epsilon^{BC}$ is the Levi-Civita symbol in the two-dimensional projected space. We call the latter $\cVT{}_A(\eta, \bm{k})$ a curl vector. The auto-correlations of these two vectors are defined by
\bae{
\Braket{\VT{}^{A} (\eta, \bm{k}) \VT{}_{A} (\eta^\prime, \bm{k}^\prime)} & = \left(2\pi\right)^3\DD  (\bm{k}+\bm{k}^\prime) F^{\mathrm{T}}_{VV} (\muk, \eta, \eta^\prime, k) ~, 
\\
\Braket{\cVT{}^{A} (\eta, \bm{k}) \cVT{}_{A} (\eta^\prime, \bm{k}^\prime)} & = \left(2\pi\right)^3\DD  (\bm{k}+\bm{k}^\prime) F^{\mathrm{T}}_{\cV \cV} (\muk, \eta, \eta^\prime, k) ~.
}
It is immediate to verify that the two functions $F^{\mathrm{T}}_{VV}$ and $F^{\mathrm{T}}_{\cV \cV}$ are identical as 
\bae{
F^{\mathrm{T}}_{VV} (\muk, \eta, \eta^\prime, k) ~
&\begin{aligned}
= ~& F^{\mathrm{T}}_{\cV \cV} (\muk, \eta, \eta^\prime, k) ~&
\end{aligned}
 \nonumber\\
&\begin{aligned}
= ~& \frac{1}{16} k^2 \left(1-\muk^2\right) \left(1+6\muk^2+\muk^4\right)
\left\{c_{+2}^2 P^{(+2)} (\eta, \eta^\prime, k) + c_{-2}^2 P^{(-2)} (\eta, \eta^\prime, k)\right\}
\\
&
+ \frac14 k^2 \left(1 - \muk^2\right)\left(1 - \muk^4\right)
\left\{c_{+1}^2 P^{(+1)} (\eta, \eta^\prime, k) + c_{-1}^2 P^{(-1)} (\eta, \eta^\prime, k)\right\}
\\
& + \frac18 k^2\left(1 - \muk^2\right)^3 c_{b}^2 P^{(b)}(\eta, \eta^\prime, k) 
+ \frac14 k^2\left(1 - \muk^2\right)^3 c_{\ell}^2 P^{(\ell)}(\eta, \eta^\prime, k)
~,
\end{aligned}
}
which is parity-even as evidenced by eqs.~\eqref{eq. vector without trace} and \eqref{eq. curl vector without trace}. 
Note that the $\muk$-dependence for the helicity-$2$ mode corresponds to eq.~(3.11) of ref.~\cite{Okumura:2024xnd}.
Besides, by comparing with eq.~\eqref{eq. F function tensor}, we see that the following holds true:
\bae{
\Braket{\VT{}^{A} (\eta, \bm{k}) \VT{}_{A} (\eta^\prime, \bm{k}^\prime)} = \frac12 k^2 \left(1-\muk^2\right)\Braket{\hhT{}^{AB}(\eta, \bm{k}) \hhT{}_{AB} (\eta^\prime, \bm{k}^\prime)} ~.
}

In order to see parity-violating signals, it is necessary to consider correlations that mix diagonal and off-diagonal components of the matrix~\eqref{eq. matrix of traceless projected tensor}.
The cross-correlation between the vector and curl vector accomplishes this, whose power spectrum is defined as
\bae{\label{eq. correlation function vector cross}
  \Braket{\cVT{}^{A} (\eta, \bm{k}) \VT{}_{A} (\eta^\prime, \bm{k}^\prime)} & = \left(2\pi\right)^3\DD  (\bm{k}+\bm{k}^\prime) F^{\mathrm{T}}_{\cV V} (\muk, \eta, \eta^\prime, k) ~,
}
with
\beae{\label{eq. F function vector cross}
  F^{\mathrm{T}}_{\cV V} =~& -\frac{i}{4} k^2 \muk \left(1-\muk^4\right) \left\{c_{+2}^2 P^{(+2)} (\eta, \eta^\prime, k) - c_{-2}^2 P^{(-2)} (\eta, \eta^\prime, k)\right\}
  \\
  & 
  -\frac{i}{2} k^2 \muk \left(1-\muk^2\right)^2 \left\{c_{+1}^2 P^{(+1)} (\eta, \eta^\prime, k) - c_{-1}^2 P^{(-1)} (\eta, \eta^\prime, k)\right\}
  ~.
}
The $\muk$-dependence for the helicity-$2$ mode coincides with eq.~(3.14) of ref.~\cite{Okumura:2024xnd}.
We note that the same expression~\eqref{eq. F function vector cross} can be obtained if we define two vectors without extracting trace from the projected tensor field, which means
\bae{
\Braket{\cVT{}^{A} (\eta, \bm{k}) \VT{}_{A} (\eta^\prime, \bm{k}^\prime)} & = \Braket{\cV^{A} (\eta, \bm{k}) V_{A} (\eta^\prime, \bm{k}^\prime)} ~,
}
with
\bae{
  V_{A} (\eta, \bm{k}) \coloneqq i k^{B} \hat{h}_{BA}(\eta, \bm{k})  ~, \quad
  \cV_A(\eta, \bm{k})  \coloneqq i \epsilon^{BC}k_{B} \hat{h}_{CA}(\eta, \bm{k}) ~.
}
\subsection{Projected scalar field}\label{sec. Projected scalar perturbation}
Let us finally look at correlations of scalar quantities.
Using the traceless property of $\hhT{}_{AB}$, one can define two scalar quantities as
\bae{
  \ST (\eta, \bm{k}) & \coloneqq - k^{A} k^{B} \hhT{}_{AB}(\eta, \bm{k}) = k^2  \sin^2{\theta_k} \hhT{}_{xx} ~,
  \\
  \cST(\eta, \bm{k}) & \coloneqq - \epsilon^{AC} k_{C} k^{B} \hhT{}_{AB}(\eta, \bm{k}) =-k^2  \sin^2{\theta_k} \hhT{}_{xy} ~,
}
where we call the second a curl scalar. The definitions of the two scalars imply that their auto-correlations are parity-even and their cross-correlation can lead to parity-violating signals.
The auto-correlation of the scalar becomes
\bae{
\Braket{\ST (\eta, \bm{k}) \ST (\eta^\prime, \bm{k}^\prime)} = \left(2\pi\right)^3\DD  (\bm{k}+\bm{k}^\prime) F^{\mathrm{T}}_{S S} (\muk, \eta, \eta^\prime, k) ~,
}
with
\beae{
F^{\mathrm{T}}_{SS} (\muk, \eta, \eta^\prime, k) = ~& 
\frac{1}{16} k^4 \left(1-\muk^4\right)^2
\left\{c_{+2}^2 P^{(+2)} (\eta, \eta^\prime, k) + c_{-2}^2 P^{(-2)} (\eta, \eta^\prime, k)\right\}
\\
&
+ \frac14 k^4\muk^2 \left(1 - \muk^2\right)^3
\left\{c_{+1}^2 P^{(+1)} (\eta, \eta^\prime, k) + c_{-1}^2 P^{(-1)} (\eta, \eta^\prime, k)\right\}
\\
& + \frac18 k^4\left(1 - \muk^2\right)^4 c_{b}^2 P^{(b)}(\eta, \eta^\prime, k)
+ \frac14 k^4\left(1 - \muk^2\right)^4 c_{\ell}^2 P^{(\ell)}(\eta, \eta^\prime, k)
~,
}
while one of the curl scalar is given by
\bae{
\Braket{\cST (\eta, \bm{k}) \cST (\eta^\prime, \bm{k}^\prime)} = \left(2\pi\right)^3\DD  (\bm{k}+\bm{k}^\prime) F^{\mathrm{T}}_{\cS\cS} (\muk, \eta, \eta^\prime, k) ~,
}
with
\beae{
F^{\mathrm{T}}_{\cS\cS} (\muk, \eta, \eta^\prime, k) = ~& 
\frac{1}{4} k^4 \muk^2 \left(1-\muk^2\right)^2
\left\{c_{+2}^2 P^{(+2)} (\eta, \eta^\prime, k) + c_{-2}^2 P^{(-2)} (\eta, \eta^\prime, k)\right\}
\\
& 
+ \frac14 k^4 \left(1 - \muk^2\right)^3
\left\{c_{+1}^2 P^{(+1)} (\eta, \eta^\prime, k) + c_{-1}^2 P^{(-1)} (\eta, \eta^\prime, k)\right\}
~.
}
The $\muk$-dependence for the helicity-$2$ mode agrees with eq.~(3.13) of ref.~\cite{Okumura:2024xnd}.
Finally, the cross-correlation of the standard scalar and curl scalar is expressed as
\bae{
\Braket{\cST (\eta, \bm{k}) \ST (\eta^\prime, \bm{k}^\prime)} = \left(2\pi\right)^3\DD (\bm{k}+\bm{k}^\prime) F^{\mathrm{T}}_{\cS S} (\muk, \eta, \eta^\prime, k) ~,
}
where the explicit form of the function $F^{\mathrm{T}}_{\cS S}$ is given by
\beae{
F^{\mathrm{T}}_{\cS S} (\muk, \eta, \eta^\prime, k) = ~& -\frac{i}{8} k^4 \muk \left(1-\muk^2\right)^2\left(1+\muk^2\right) \left\{c_{+2}^2 P^{(+2)} (\eta, \eta^\prime, k) - c_{-2}^2 P^{(-2)} (\eta, \eta^\prime, k)\right\}
\\
& -\frac{i}{4} k^4 \muk \left(1-\muk^2\right)^3 \left\{c_{+1}^2 P^{(+1)} (\eta, \eta^\prime, k) - c_{-1}^2 P^{(-1)} (\eta, \eta^\prime, k)\right\}
~.
}
We note that this function is related to the vector cross-correlation as 
\bae{
  F^{\mathrm{T}}_{\cS S} = \frac12 k^2 \left(1-\muk^2\right) F^{\mathrm{T}}_{\cV V} ~.
}
\section{The overlap reduction function}\label{sec. ORF}
We turn to the properties of the two-point correlation functions. We provide our definition of the \ac{ORF} in section~\ref{sec. ORF general} and present three concrete examples in section~\ref{sec. Computing the ORF}.
\subsection{Definition}\label{sec. ORF general}
Letting $X$ and $Y$ be rank-$n$ tensors such as $\hhT{}_{AB}$ and $\VT{}_{A}$, the two-point correlation function may be written as
\bae{\label{eq. two-point}
\Braket{X (\eta, \bm{x}) Y (\eta^\prime, \bm{x}^\prime)} = \int \frac{k^2 \dd k}{(2\pi)^3}\int\dd\Omega_k \left[k^{2(2-n)} \tilde{F}_{XY} (\muk, \eta, \eta^\prime, k) \ee^{i\bm{k}\cdot \left(\bm{x}-\bm{x}^\prime\right)}\right] ~,
}
where $\tilde{F}_{XY} \coloneqq F_{XY}/k^{2(2-n)} \propto k^0$ and $\dd\Omega_k$ denotes the angular element in the Fourier space.\footnote{This expression is no longer true for correlations where the trace of the tensor $g^{AB}h_{AB}$ is involved.} 
Recalling that the power spectrum with helicity $\lambda$ is given in eq.~\eqref{eq. Power dynamics}, we can write
\bae{
\tilde{F}_{XY} (\muk, \eta, \eta^\prime, k) 
& = \sum_\lambda f^{(\lambda)} (\muk) P^{(\lambda)} (\eta, \eta^\prime, k) ~,
\nonumber \\
& = \frac{a^2(\eta_\ini)}{a(\eta)a(\eta^\prime)} \sum_\lambda \cos\left[c_{\lambda} k \left(\eta-\eta^\prime\right)\right] f^{(\lambda)} (\muk) P^{(\lambda)}_{\mathrm{ini}}(k) ~.
}
Here, the function $f^{(\lambda)} (\muk)$ depends on a specific choice of correlations. 

We restrict ourselves to the case where all helicity modes have the same spectral characteristics allowing different amplitudes. Under this assumption, we can parametrize as
\bae{
P^{(\lambda)}_{\mathrm{ini}}(k) = N^{(\lambda)} P_{h, \mathrm{ini}}(k) ~,
}
with a condition
\bae{
  \sum_{\lambda} N^{(\lambda)} = 1  ~.
}
This assumption enables us to factorize the two-point function as
\bae{\label{eq. ORF in two-point}
\Braket{X (\eta, \bm{x}) Y (\eta^\prime, \bm{x}^\prime)} = \frac{a^2 (\eta_\ini)}{a(\eta)a(\eta^\prime)} \int \frac{k^{2(3-n)} \dd k}{(2\pi)^3} \Gamma_{XY}  P_{h, \mathrm{ini}}(k) ~,
}
where $\Gamma_{XY}$ is given by
\bae{\label{eq. def ORF}
\Gamma_{XY} \coloneqq \int\dd\Omega_k \hGamma_{XY}\ee^{i\bm{k}\cdot \left(\bm{x}-\bm{x}^\prime\right)} ~,
}
with 
\bae{\label{eq. hGamma}
\hGamma_{XY} \coloneqq \sum_\lambda \cos\left[c_{\lambda} k \left(\eta-\eta^\prime\right)\right] f^{(\lambda)} (\muk) N^{(\lambda)} ~.
}
We call the kernel $\Gamma_{XY}$ an \ac{ORF} following the terminology used in the \ac{PTA}.

In order to analyze eq.~\eqref{eq. def ORF}, we adopt the flat-sky approximation and choose the coordinate system in the real space as
\bae{
  x^\prime{}^\mu = \left( \eta^\prime, 0, 0, r^\prime \right) ~, \quad 
  x{}^\mu = \left( \eta, r \sin\theta \cos\phi, r\sin\theta \sin\phi, r\cos\theta \right) ~,
}
meaning that we take the \ac{LOS} direction to point toward a galaxy at $\bm{x}^\prime$ and consider that another galaxy is located at a point $\bm{x}$. The wavevector can point to arbitrary directions, so that we take
\bae{\label{eq. coord wave}
  \bm{k} = \left( k \sin\theta_k \cos\phi_k, k \sin\theta_k \sin\phi_k, k\cos\theta_k \right) ~.
}
The angular integral of eq.~\eqref{eq. def ORF} can be performed by means of the spherical harmonics expansion.\footnote{In ref.~\cite{Okumura:2024xnd}, the angular integral is directly integrated using the Bessel function. This method is explained in appendix~\ref{sec. the angular integral}.} Letting $\bm{R}=\bm{x}-\bm{x}^\prime$ be a separation vector from $\bm{x}^\prime$ to $\bm{x}$ and using the spherical Bessel function $j_\ell$ and the spherical harmonics $Y_{\ell m}$, we can write
\bae{\label{eq. ylm exp}
\ee^{i\bm{k}\cdot \bm{R}} = 4\pi \sum_{\ell =0}^{\infty}\sum_{m=-\ell}^{\ell} i^\ell Y^\ast_{\ell m}(\theta_k, \phi_k) j_\ell (k R) Y_{\ell m}(\theta_R, \phi_R) ~,
}
where $R\coloneqq |\bm{R}|$, and $\ell$ and $m$ are degree and order of the spherical harmonics, respectively. We remark that $\theta_R$ and $\phi_R$ are angles between $\bm{x}^\prime$ and $\bm{R}$.
Now, the model-dependent part in the \ac{ORF}~\eqref{eq. def ORF} can also be expanded by the spherical harmonics as
\bae{
f^{(\lambda)} (\muk) N^{(\lambda)} = \sum_{\ell^\prime=0}^{\infty} a^{(\lambda)}_{\ell^\prime 0} Y_{\ell^\prime 0}(\theta_k)
~,
}
where $a^{(\lambda)}_{\ell^\prime 0}$ denote model-dependent coefficients. We have imposed a condition that the order is zero because the coordinate system have been chosen so that the azimuthal angle $\phi_k$ is zero (see eq.~\eqref{eq. coord wave}). It is now clear that the angular integral can be performed using the property of the spherical harmonics
\bae{\label{eq. property of harmonics}
\int \dd\Omega_k Y^\ast_{\ell m}(\theta_k, \phi_k) Y_{\ell^\prime m^\prime}(\theta_k, \phi_k) = \delta_{\ell \ell^\prime} \delta_{m m^\prime} ~.
}
We finally note that the \ac{ORF} is a function of the distances and the angle from an observer, while, after using the property~\eqref{eq. property of harmonics}, we are left with the separation between two galaxies $R$ and the angle between the \ac{LOS} and the separation vector $\theta_R$. Defining 
\bae{
  \Delta r \coloneqq r^\prime - r = \eta - \eta^\prime ~,
}
which is bounded as $\Delta r > -r$, the separation distance $R$ can be rewritten as
\bae{
R = \sqrt{r^2+\left(r^\prime\right)^2-2 rr^\prime \cos{\theta}} 
= r \sqrt{1+\left(1 + \frac{k\Delta r}{k r}\right)^2-2\left(1 + \frac{k\Delta r}{k r}\right)\cos{\theta}} ~,
}
and the angles are related as
\bae{
\theta_R = \theta + \arccos{\frac{1-\left(1+\frac{k\Delta r}{kr}\right)\cos{\theta}}{\sqrt{1+\left(1+ \frac{k\Delta r}{k r}\right)^2 -2 \left(1+\frac{k\Delta r}{k r}\right)\cos{\theta}}}} ~.
}

\subsection{Computing the ORF: examples}\label{sec. Computing the ORF}
As representative examples, let us investigate the ORFs of the projected tensor, vector, and scalar fields in the presence of the helicity-one and helicity-zero modes. While the power spectrum of the longitudinal mode is significantly larger than that of the other modes, we treat $N^{(\lambda)}$ as free parameters in this work for illustration purposes. We assume that the longitudinal mode corresponding to the gravitational potential is constant in time and the helicity-two modes propagate at the speed of light. These assumptions would be reasonable considering current cosmological observations and the GW170817 event~\cite{LIGOScientific:2017zic}.
\subsubsection{Auto-correlation of the traceless tensor field}
Let us first consider the auto-correlation of the traceless tensor field assuming that $\cv\coloneqq c_{+1}=c_{-1}$.
The function $\hGamma$, defined in eq.~\eqref{eq. hGamma}, is a function of the directional cosine $\muk$ and can be written in terms of the spherical harmonics as 
\bae{\label{eq. ORF hThT}
\hGamma_{\hhT\hhT} = \sum_{i=0,2,4} C^{\hhT \hhT}_i Y_{i 0}(\theta_k) ~,
}
where coefficients $C^{\hhT \hhT}_i$ are explicitly given by
\bae{
  C^{\hhT \hhT}_0 & = \frac{4\sqrt{\pi}}{15}\left[3 N^{(\mathrm{T})}\cos\left(k\Delta r\right)
  + 3 \sum_{\lambda = \pm 1} c_\lambda^2 N^{(\lambda)}\cos\left(c_\lambda k\Delta r\right)
  + c_b^2 N^{(b)} \cos\left(c_{b} k\Delta r\right)
  + 2 N^{(\ell)}
  \right] ~, 
  \\
  C^{\hhT \hhT}_2 & = \frac{4}{21}\sqrt{\frac{\pi}{5}}\left[
  6 N^{(\mathrm{T})} \cos\left(k\Delta r\right)
  - 3 \sum_{\lambda = \pm 1} c_\lambda^2 N^{(\lambda)}\cos\left(c_\lambda k\Delta r\right)
  - 2 c_b^2 N^{(b)} \cos\left(c_{b} k\Delta r\right)
  - 4 N^{(\ell)}
  \right] ~, 
\\
  C^{\hhT \hhT}_4 & = \frac{2\sqrt{\pi}}{105}\left[
  N^{(\mathrm{T})}\cos\left(k\Delta r\right)
    - 4 \sum_{\lambda = \pm 1} c_\lambda^2 N^{(\lambda)}\cos\left(c_\lambda k\Delta r\right)
    + 2 c_b^2 N^{(b)} \cos\left(c_{b} k\Delta r\right)
    + 4 N^{(\ell)}
    \right] ~,
}
where we defined $N^{(\mathrm{T})} \coloneqq N^{(+2)} + N^{(-2)}$.
Recalling the property of the spherical harmonics~\eqref{eq. property of harmonics}, it is immediate to see that the \ac{ORF} takes the form
\bae{\label{eq. ORF YLM}
\Gamma_{\hhT\hhT} =
  \sum_{\ell=0,2,4} i^\ell C^{\hhT \hhT}_\ell Y_{\ell 0}(\theta_R) j_\ell (k R) ~.
}

\begin{figure}
  \centering
  \includegraphics[width=.8\hsize]{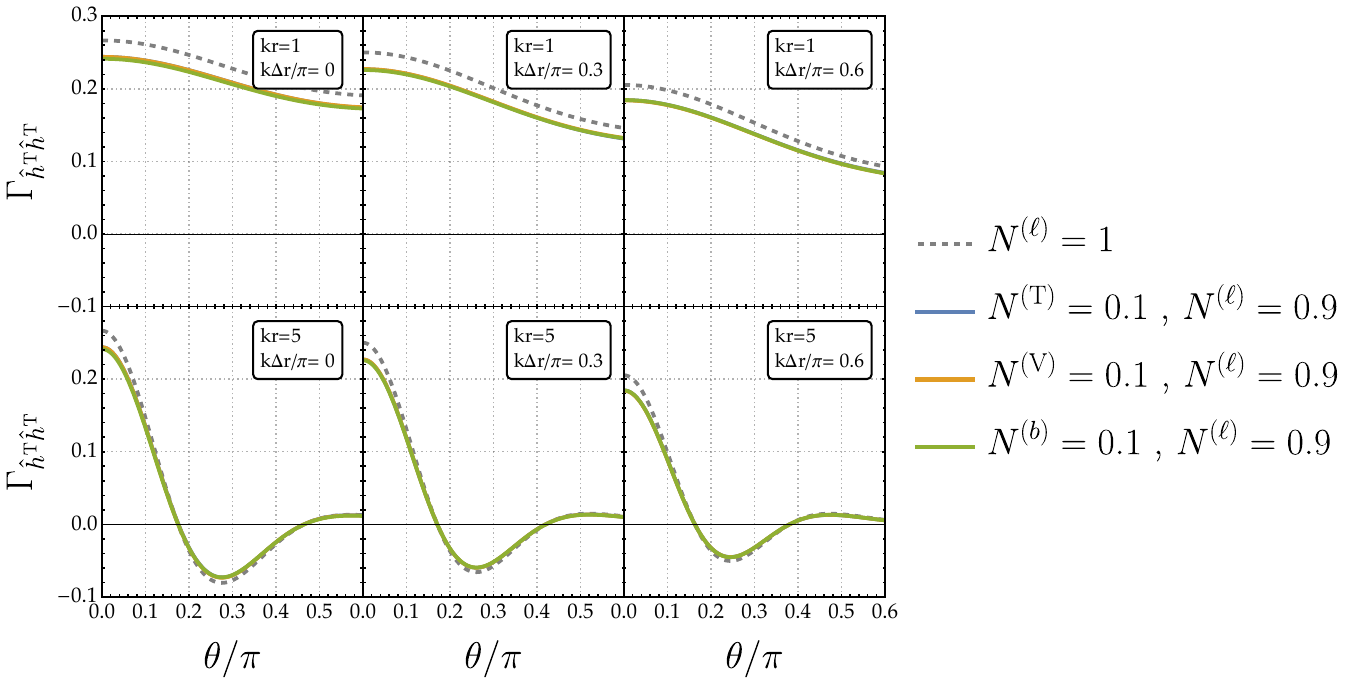}
  \caption{
    The auto-correlation of the projected traceless tensor field with $c_\lambda=1$. The gray dashed line corresponds to the case where only the longitudinal component is present, which can be regarded as a fiducial line. Other lines are depicted by assuming that $10\%$ of the total power is distributed to other modes.
  }
  \label{Fig. plothhTn}
\end{figure}

Figure~\ref{Fig. plothhTn} shows the \ac{ORF} of the projected traceless tensor \eqref{eq. ORF YLM} with propagation speed $c_\lambda=1$.
The gray dashed line is plotted by imposing $N^{(\ell)}=1$, which can be regarded as a fiducial line because this auto-correlation is dominated by the longitudinal mode. The other lines are depicted by assuming that the tensor, vector, or breathing mode carries $10\%$ of the total power. As can be seen, as long as the longitudinal mode predominantly contributes to the total power, the amplitude and shape of the \ac{ORF} do not change significantly, which makes it difficult to probe GW polarizations with this correlation.
\subsubsection{Cross-correlation of vector fields}
Let us next investigate the \ac{ORF} of the vector cross-correlation which is suitable for probing parity violation.
As can be seen in eq.~\eqref{eq. F function vector cross}, the function $\hGamma$ in this case is an odd polynomial of $\muk$, so that it can be expanded by the spherical harmonics with odd degree
\bae{\label{eq. ORF cVTVT}
\hGamma_{\cVT\VT} = \sum_{\ell=1,3,5} C^{\cVT\VT}_\ell Y_{\ell 0}(\theta_k) ~,
}
where the coefficients $C^{\cVT\VT}_\ell$ are explicitly given by
\bae{
  C^{\cVT\VT}_1 & = - i \frac{2}{35}\sqrt{\frac{\pi}{3}}\left[5 \widetilde{N}^{(\mathrm{T})} \cos\left(k\Delta r\right)
  + 4 c_{+1}^2 N^{(+1)} \cos\left(c_{+1} k\Delta r\right)
  - 4 c_{-1}^2 N^{(-1)} \cos\left(c_{-1} k\Delta r\right)
  \right] ~, 
  \\
  C^{\cVT\VT}_3 & = i \frac{2}{45}\sqrt{\frac{\pi}{7}}\left[
  5 \widetilde{N}^{(\mathrm{T})} \cos\left(k\Delta r\right)
  + 8 c_{+1}^2 N^{(+1)} \cos\left(c_{+1} k\Delta r\right)
  - 8 c_{-1}^2 N^{(-1)} \cos\left(c_{-1} k\Delta r\right)
  \right] ~, 
  \\
  C^{\cVT\VT}_5 & = i \frac{4}{63}\sqrt{\frac{\pi}{11}}\left[
    \widetilde{N}^{(\mathrm{T})} \cos\left(k\Delta r\right)
    -2 c_{+1}^2 N^{(+1)} \cos\left(c_{+1} k\Delta r\right)
    +2 c_{-1}^2 N^{(-1)} \cos\left(c_{-1} k\Delta r\right)
    \right] ~,
}
with $\widetilde{N}^{(\mathrm{T})} \coloneqq N^{(+2)} - N^{(-2)}$.
As in the previous subsection, the \ac{ORF} is expressed as
\bae{\label{eq. ORF cVV}
\Gamma_{\cVT\VT} =
  \sum_{\ell=1,3,5} i^{\ell} C^{\cVT\VT}_\ell Y_{\ell 0}(\theta_R) j_\ell (k R) ~.
}

\begin{figure}
  \centering
  \includegraphics[width=.8\hsize]{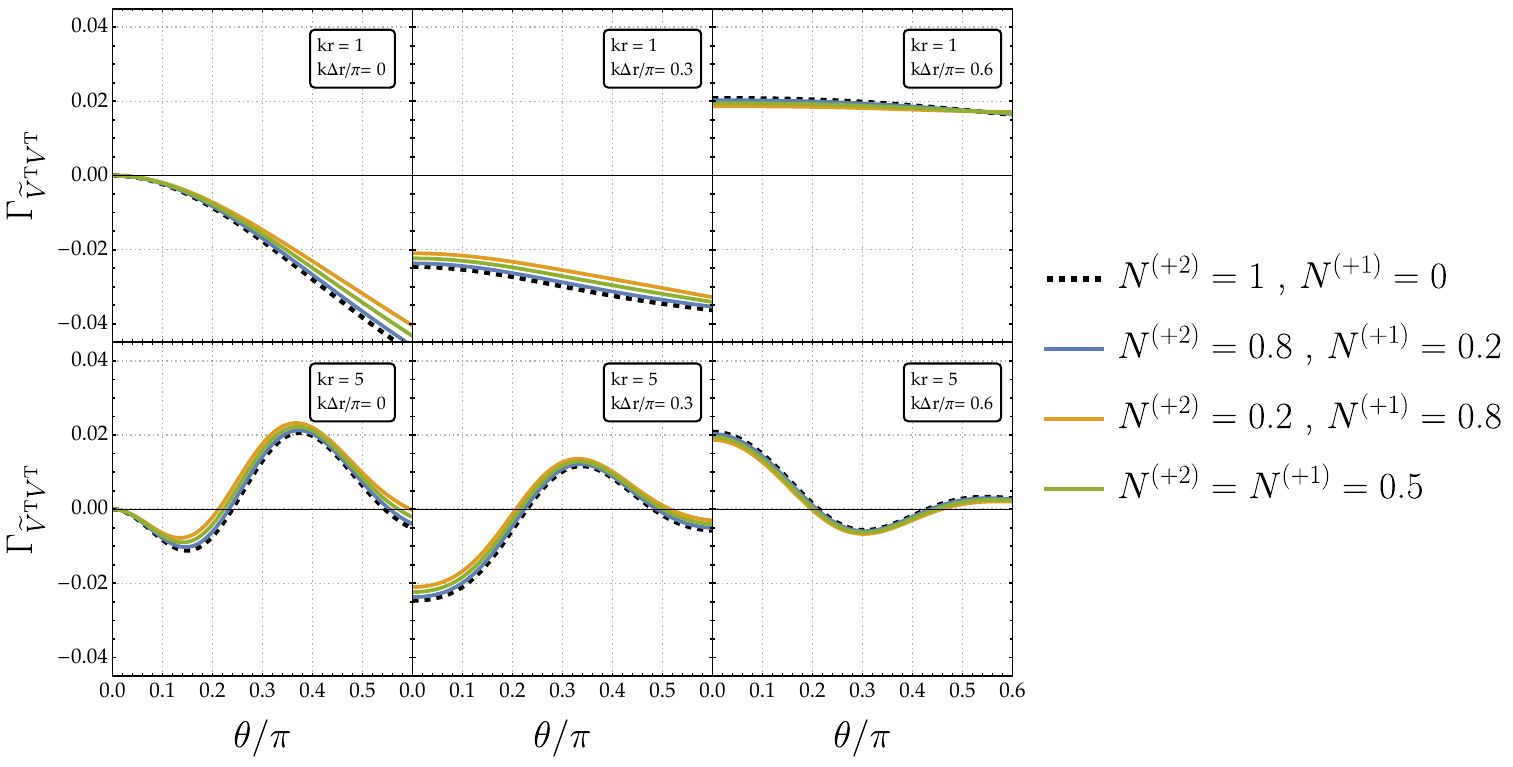}
  \caption{
  The cross-correlation of the vector and curl vector assuming that the helicity-two and helicity-one modes are maximally chiral, that is $P^{(-2)}=P^{(-1)}=0$. We also assume that the speed of the helicity-one wave equals to the speed of light. The black dashed line is depicted without the vector mode. 
  }
  \label{Fig. PlotcVVT}
\end{figure}

\begin{figure}
  \centering
  \includegraphics[width=.85\hsize]{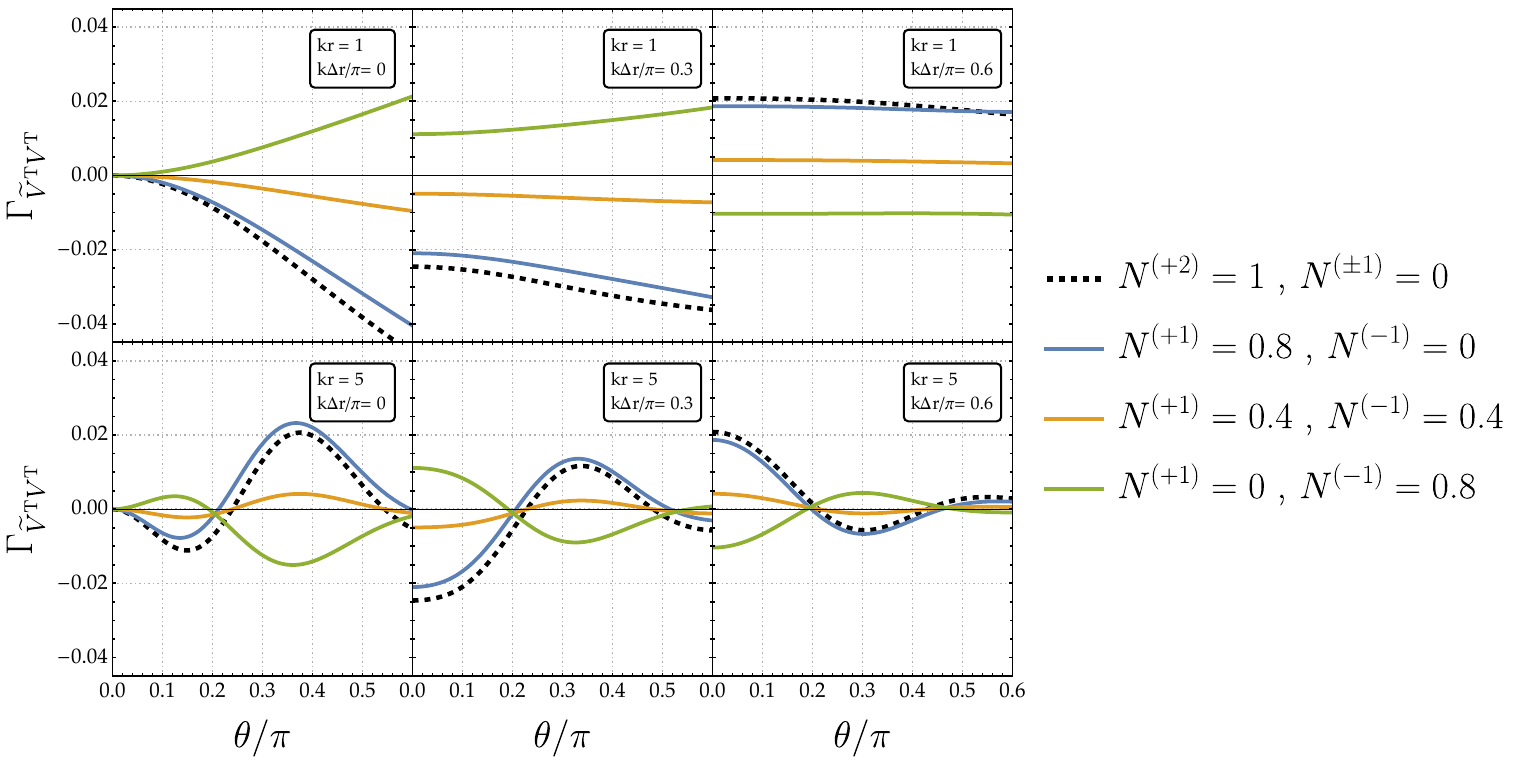}
  \caption{
    The cross-correlation of the vector and curl vector without assuming the maximally chirality for helicity-one waves. We assume that the speed of the helicity-one waves equals to the speed of light.
  }
  \label{Fig. PlotcVVThalf}
\end{figure}

Figure~\ref{Fig. PlotcVVT} shows the \ac{ORF} of the vector cross-correlation assuming a maximal chirality for the helicity-two and helicity-one modes, namely, $P^{-2} = P^{-1} = 0$.
We also assume that the helicity-one wave propagates at the speed of light. The black dashed line shows the \ac{ORF} in the case when only the wave with helicity $+2$ is present, while other lines contain some amount of the vector wave with helicity $+1$. One sees that, while the amplitude slightly depends on the amount of the helicity-one mode, the shape of the \ac{ORF} remains almost unchanged due to the maximal chirality.
If there exists a certain amount of the vector wave with helicity $-1$, the sign of the \ac{ORF} can be different from the fiducial line, which can be seen in figure~\ref{Fig. PlotcVVThalf}.

\begin{figure}[bt]
  \centering
  \includegraphics[width=.75\hsize]{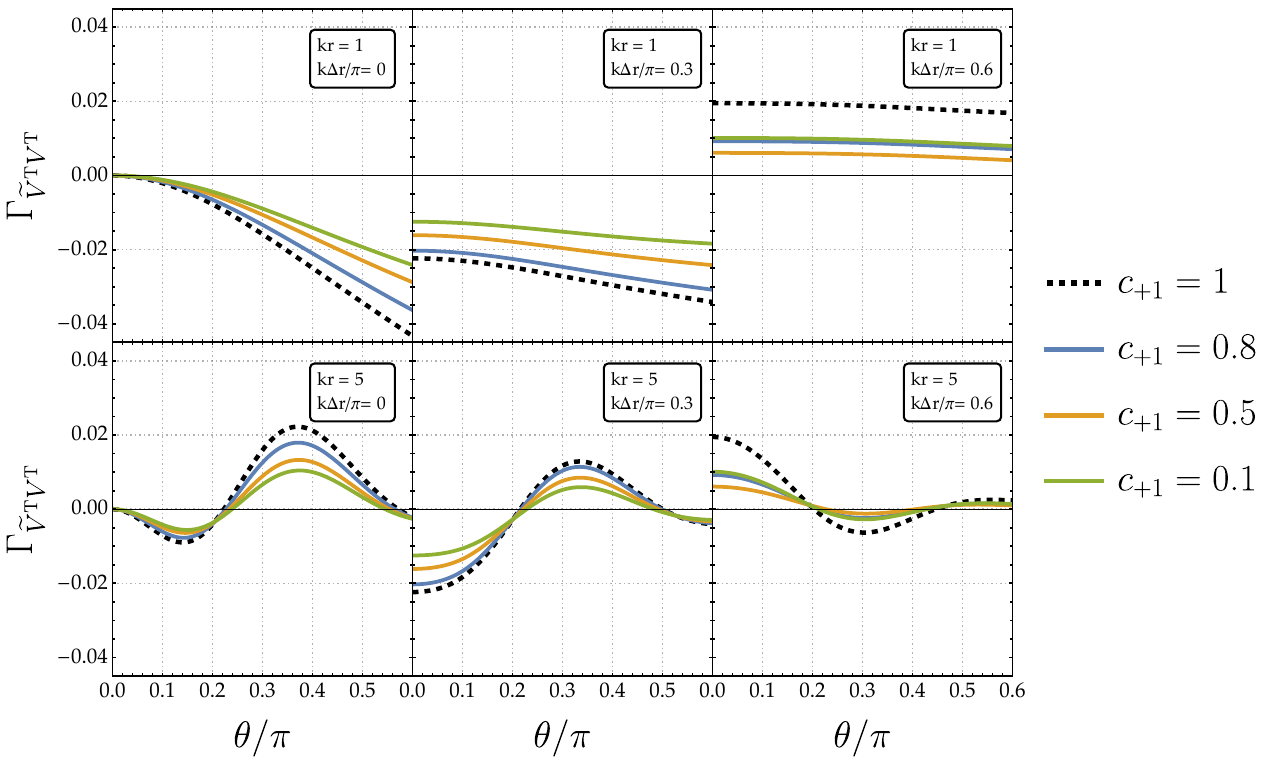}
  \caption{
    Dependence of the speed of the helicity-one wave on the cross-correlation of the vector and curl vector assuming a maximal chirality for the helicity-two and helicity-one modes. The distribution of the power is fixed to be $N^{(+2)}=N^{(+1)}=0.5$, which corresponds to the green line in figure~\ref{Fig. PlotcVVT}.
  }
  \label{Fig. PlotcVVTcs}
\end{figure}

When the propagation speed differs from mode to mode, the shape of the \ac{ORF} can be different. Figure~\ref{Fig. PlotcVVTcs} shows the dependence of the propagation speed of the helicity-one wave on the vector cross-correlation. 
We can see that, if the helicity-one mode propagates slower than light, the \ac{ORF} exhibits different behaviors, which come from an effective suppression of the power spectra by the propagating speed and different oscillatory functions $\cos\left(c_{\lambda} k\Delta r\right)$. Thus, if parity violation exists and the propagation speeds of the modes differ, upcoming observations of the large-scale structure may provide insights into the underlying physics of the primordial universe.
\subsubsection{Auto-correlation of the curl scalar}
Let us finally study the auto-correlation of the curl scalar $\cST$, which is free of the contribution of the longitudinal mode. The function $\hGamma$ for this correlation can also be expanded by the spherical harmonics with even degree as
\bae{\label{eq. ORF csTcsT}
\hGamma_{\cST\cST} = \sum_{\ell=0,2,4,6} C^{\cST\cST}_\ell Y_{\ell 0}(\theta_k) ~,
}
where their coefficients $C^{\cST\cST}_\ell$ are explicitly given by
\bae{
  C^{\cST\cST}_0 & = \frac{4\sqrt{\pi}}{105}\left[N^{(\mathrm{T})}\cos\left(k\Delta r\right)
  + 6 \sum_{\lambda = \pm 1} c_\lambda^2 N^{(\lambda)}\cos\left(c_\lambda k\Delta r\right)
  \right] ~, 
  \\
  C^{\cST\cST}_2 & = -\frac{8}{21}\sqrt{\frac{\pi}{5}}\left[
    \sum_{\lambda = \pm 1} c_\lambda^2 N^{(\lambda)}\cos\left(c_\lambda k\Delta r\right)
  \right] ~, 
\\
  C^{\cST\cST}_4 & = - \frac{4\sqrt{\pi}}{1155}\left[
  7 N^{(\mathrm{T})}\cos\left(k\Delta r\right)
    - 18 \sum_{\lambda = \pm 1} c_\lambda^2 N^{(\lambda)}\cos\left(c_\lambda k\Delta r\right)
    \right] ~,
\\
C^{\cST\cST}_6 & = \frac{8}{231}\sqrt{\frac{\pi}{13}}\left[
  N^{(\mathrm{T})}\cos\left(k\Delta r\right)
  - \sum_{\lambda = \pm 1} c_\lambda^2 N^{(\lambda)}\cos\left(c_\lambda k\Delta r\right)
    \right] ~.
}
Then, using the property of the spherical harmonics, the \ac{ORF} takes the form
\bae{\label{eq. ORF cScS}
  \Gamma_{\cST\cST} =
  \sum_{\ell=0,2,4,6} i^{\ell} C^{\cST\cST}_\ell Y_{\ell 0}(\theta_R) j_\ell (k R) ~.
}

\begin{figure}
  \centering
  \includegraphics[width=.8\hsize]{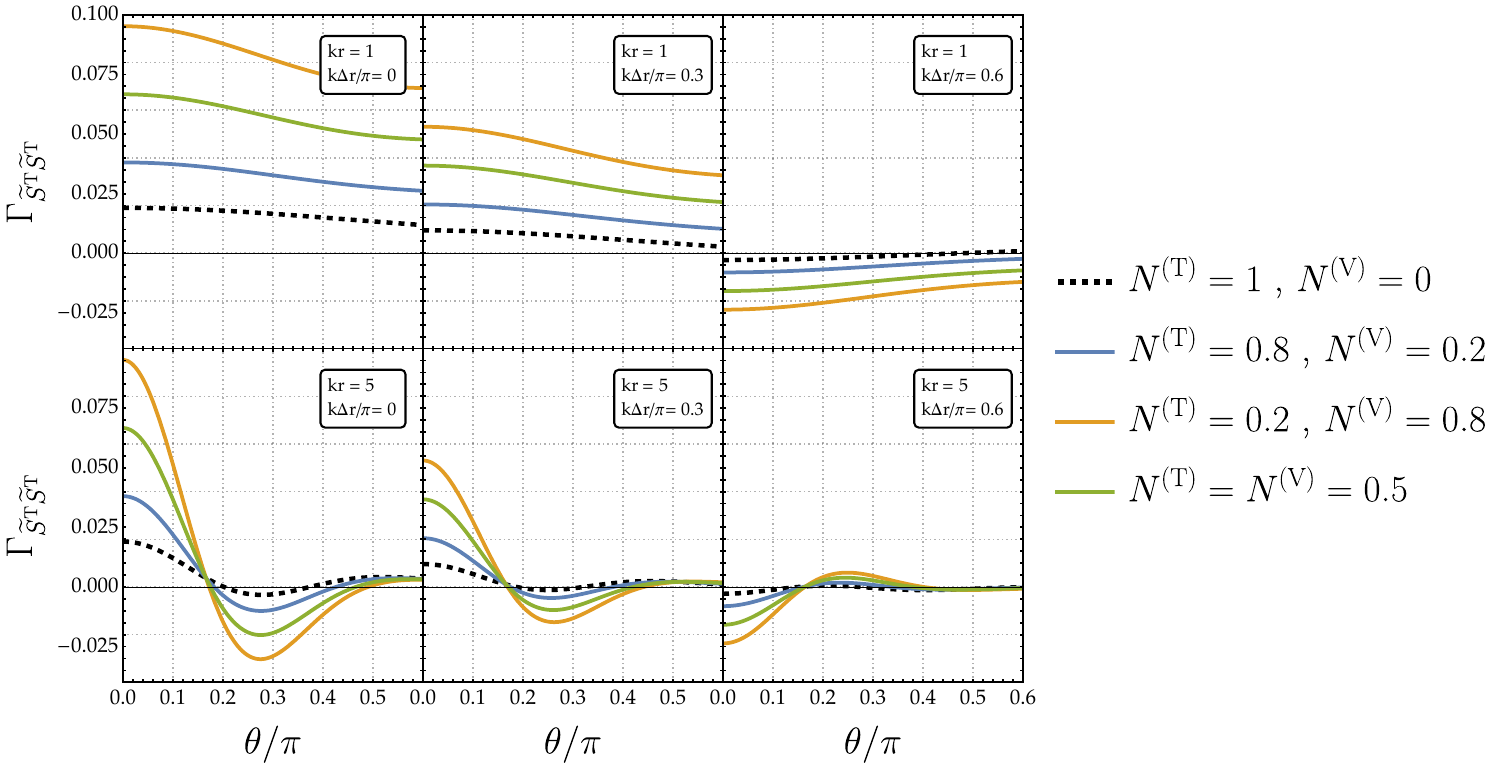}
  \caption{
  The auto-correlation of the curl scalar assuming that the helicity-two and helicity-one modes propagate with the speed of light. The black dashed line is drawn by neglecting the vector mode.
  }
  \label{Fig. PlotcSTcsT}
\end{figure}

\begin{figure}[bt]
  \centering
  \includegraphics[width=.75\hsize]{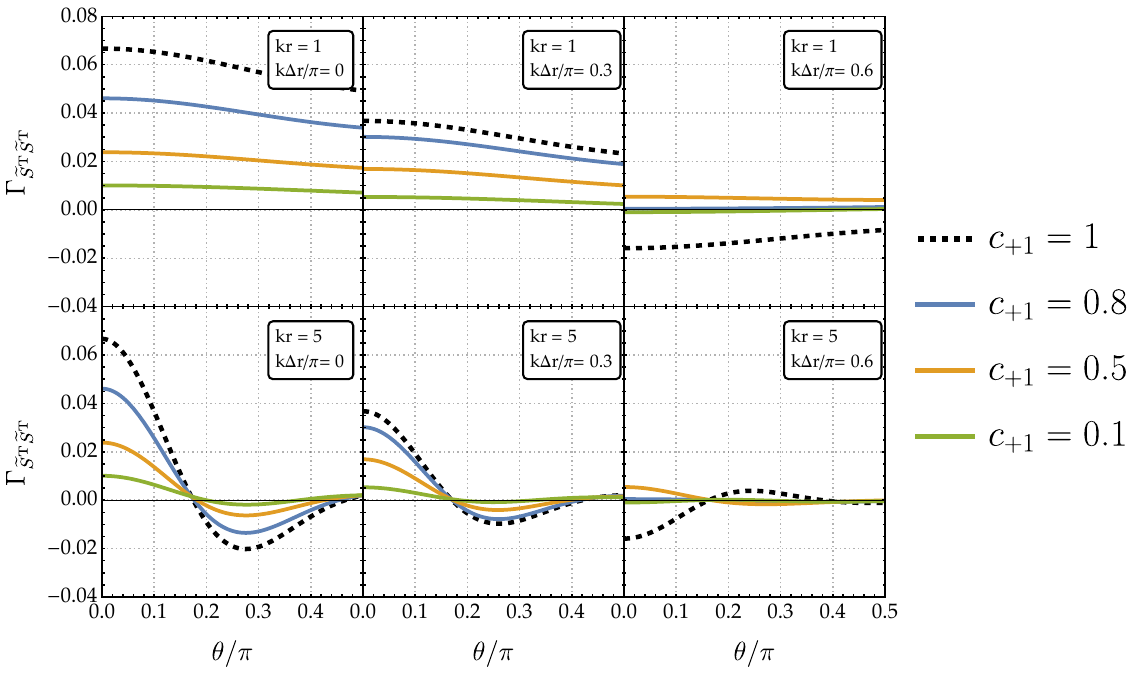}
  \caption{
    Dependence of the speed of the helicity-one wave on the auto-correlation of the curl scalar. The distribution of the power is fixed to be $N^{(\mathrm{T})}=N^{(\mathrm{V})}=0.5$.
  }
  \label{Fig. PlotcSTcsTcs}
\end{figure}

Figure~\ref{Fig. PlotcSTcsT} shows the dependence of the \ac{ORF} on the additional vector modes assuming $\cv= c_{+1}=c_{-1}$. 
The black dashed line corresponds to the case where only the helicity-two modes are present, thereby being a fiducial line.
It is clear that the amplitude of the \ac{ORF} is different depending on the amount of the vector modes. Thus, in principle, an accurate data of galaxy redshift survey enables us to verify whether or not the \ac{GWs} carry extra polarizations on top of the helicity-two modes. 
When performing template matching, again, we need to pay attention to the possibility where the propagation speeds of the vector modes can be different from the speed of light. Figure~\ref{Fig. PlotcSTcsTcs} shows the dependence of the propagation speed of the vector modes on the \ac{ORF}, indicating that the amplitude and sign of the \ac{ORF} depend on the propagating speed.
\section{Conclusions}\label{sec. Conclusion}
We have studied the three-dimensional correlation functions using the tensor field projected onto the two-dimensional space and investigated how extra polarization modes beyond \ac{GR} can change the spectra. In particular, we have focused on fields consisting of $2D$ derivatives of the traceless projected tensor field to isolate the dominant contribution to the galaxy shape distortion by the longitudinal mode. We have found that the amplitude and angular dependence of the analogue of the \ac{ORF} used in \ac{PTA} observations can vary depending on the presence of the extra polarizations, differences in their propagation speeds, and the \ac{LOS} distances to galaxies (see figure~\ref{Fig. PlotcSTcsT} and figure~\ref{Fig. PlotcSTcsTcs}). We have also studied the vector cross-correlation that can find out parity-violating signals when restricting operators constructed by the traceless projected tensor. When the extra vector modes are maximally chiral and the mode with helicity $+1$ propagates at the speed of light, the shape of the \ac{ORF} is nearly the same irrespective of the amount of the power of the vector mode, while its amplitude is slightly changed. If the mode with helicity $+1$ propagates slower than the tensor mode or if there exists a certain amount of the mode with helicity $-1$, the \ac{ORF} can exhibit different behaviors (see figures~\ref{Fig. PlotcVVT}, ~\ref{Fig. PlotcVVThalf}, and ~\ref{Fig. PlotcVVTcs}).

The \ac{IAs} contain valuable information about underlying physics of the universe and, since \ac{GWs} may leave imprints on shapes of galaxies through the tidal effect, it is possible to probe the nature of \ac{GWs} at frequencies that have not been investigated well. This work have generalized ref.~\cite{Okumura:2024xnd} to some extent in preparation for fitting modified theories of gravity with a wealth of galaxy survey data. However, we still lack theoretical foundations because we have restricted ourselves to a limited class of models with the dispersion relation $\omega_\lambda = c_\lambda k$. The form of the dispersion is modified if we consider, for instance, massive gravity or massive scalar-tensor theories. In such cases, propagation speed depends on frequency, so that the \ac{ORF} is subjected to change. Besides, we adopt the flat-sky approximation in this work to simplify the discussion. As discussed in ref.~\cite{Schmidt:2012nw}, it is important to compare with other possible sources of the IA signal such as second-order scalar contributions whose signals are smaller on large angular scales. Thus this work should be further generalized to derive full-sky expressions. We leave these studies for future research. Finally, we comment on the detectability of the imprints of \ac{GWs} on the \ac{IAs}. 
For standard \ac{GWs} from inflation, the imprints are expected to be small and the signals are much smaller than the Poissonian shape noise (see appendix~\ref{sec. EB correlation}). Although one may think that it is challenging to detect \ac{GWs} using the galaxy-shape measurements, even at the linear order, there is a possibility of generating \ac{GWs} with a higher amplitude than those originating from inflation at different scales~\cite{Gorji:2023ziy}.
Furthermore, optimizing the \ac{IAs} signal through luminosity weighting to brighter galaxies (e.g., ref.~\cite{Seljak:2009af}) may further enhance sensitivity.
\section*{Acknowledgments}
We are grateful to Romano Antonio Enea for useful discussions.
TO acknowledges support from the Taiwan National Science and Technology Council under Grants 
Nos. NSTC 112-2112-M-001-034-,
NSTC 113-2112-M-001-011- and
NSTC 114-2112-M-001-004-, and the Academia Sinica Investigator Project Grant No. AS-IV-114-M03 for the period of 2025–2029. 
This work is supported in part by JSPS KAKENHI No.~JP24K00624.

\appendix
\section{Power spectra of the projected tensor field without extracting trace}\label{sec. Power spectra of the tensor perturbation}
In section~\ref{Projection onto the celestial sphere}, we have focused on the traceless part of the projected tensor field in light of the observational difficulty in studying its trace component. Here, we list the power spectra of the projected tensor field without extracting its trace.
\subsection{Projected tensor field}
The power spectrum of the projected tensor field is given by
\bae{
\Braket{\hat{h}^{AB} (\eta, \bm{k}) \hat{h}_{AB} (\eta^\prime, \bm{k}^\prime)} = \left(2\pi\right)^3\DD (\bm{k}+\bm{k}^\prime) F_{\hat{h}\hat{h}} (\muk, \eta, \eta^\prime, k) ~,
}
with
\beae{\label{eq. tensor auto power}
F_{\hat{h}\hat{h}} (\muk, \eta, \eta^\prime, k) = ~& \frac14 \left(1+\muk^2\right)^2 
\left\{c_{+2}^2 P^{(+2)} (\eta, \eta^\prime, k) + c_{-2}^2 P^{(-2)} (\eta, \eta^\prime, k)\right\}
\\
&
+ \frac12 \left(1 + \muk^2 -2 \muk^4\right)
\left\{c_{+1}^2 P^{(+1)} (\eta, \eta^\prime, k) + c_{-1}^2 P^{(-1)} (\eta, \eta^\prime, k)\right\}
\\
& + \frac12 \left(1 + \muk^4\right) c_{b}^2 P^{(b)}(\eta, \eta^\prime, k) + \left(1 - \muk^2\right)^2 c_{\ell}^2 P^{(\ell)}(\eta, \eta^\prime, k) ~.
}
\subsection{Projected vector field}
As in section~\ref{sec. Projected vector perturbation}, one can define two different vectors out of the projected tensor field as
\bae{
V_A (\eta, \bm{k}) & \coloneqq i k^{B} \hat{h}_{BA}(\eta, \bm{k}) ~, 
\\
\quad \cV_A(\eta, \bm{k}) & \coloneqq i \epsilon^{BC}k_{B} \hat{h}_{CA}(\eta, \bm{k}) ~.
}
Their auto power spectra are defined by 
\bae{
\Braket{V^{A} (\eta, \bm{k}) V_{A} (\eta^\prime, \bm{k}^\prime)} & = \left(2\pi\right)^3\DD  (\bm{k}+\bm{k}^\prime) F_{VV} (\muk, \eta, \eta^\prime, k) ~,
\\
\Braket{\cV^{A} (\eta, \bm{k}) \cV_{A} (\eta^\prime, \bm{k}^\prime)} & = \left(2\pi\right)^3\DD  (\bm{k}+\bm{k}^\prime) F_{\cV \cV} (\muk, \eta, \eta^\prime, k)
~,
}
where the functions $F_{VV}$ and $F_{\cV \cV}$ are respectively given by
\beae{
F_{VV} (\muk, \eta, \eta^\prime, k) = ~& 
\frac14 k^2\muk^2\left(1-\muk^4\right)
\left\{c_{+2}^2 P^{(+2)} (\eta, \eta^\prime, k) + c_{-2}^2 P^{(-2)} (\eta, \eta^\prime, k)\right\}
\\
&
+ \frac14 k^2 \left(1 - \muk^2\right)^2\left(1 + 4\muk^2\right)
\left\{c_{+1}^2 P^{(+1)} (\eta, \eta^\prime, k) + c_{-1}^2 P^{(-1)} (\eta, \eta^\prime, k)\right\}
\\
& + \frac12 k^2 \muk^4\left(1 - \muk^2\right) c_{b}^2  P^{(b)}(\eta, \eta^\prime, k)
+ k^2 \left(1 - \muk^2\right)^3 c_{\ell}^2 P^{(\ell)}(\eta, \eta^\prime, k)
~,
}
and 
\beae{\label{eq. chiral vector auto power}
F_{\cV \cV} (\muk, \eta, \eta^\prime, k) = ~&
\frac14 k^2 \left(1-\muk^4\right)
\left\{c_{+2}^2 P^{(+2)} (\eta, \eta^\prime, k) + c_{-2}^2 P^{(-2)} (\eta, \eta^\prime, k)\right\}
\\
&
+ \frac14 k^2 \left(1 - \muk^2\right)^2 
\left\{c_{+1}^2 P^{(+1)} (\eta, \eta^\prime, k) + c_{-1}^2 P^{(-1)} (\eta, \eta^\prime, k)\right\}
\\
& + \frac12 k^2 \left(1 - \muk^2\right) c_{b}^2 P^{(b)}(\eta, \eta^\prime, k)
~.
}
We notice that there is no longitudinal contribution in the auto power spectrum of the curl vector $\ch_A$. 
The cross-correlation of the two different vectors becomes nonzero with chiral \ac{GWs} as
\bae{
\Braket{\cV^{A} (\eta, \bm{k}) V_{A} (\eta^\prime, \bm{k}^\prime)} = \left(2\pi\right)^3\DD  (\bm{k}+\bm{k}^\prime) F_{\cV V} (\muk, \eta, \eta^\prime, k) ~,
}
where the kernel function is 
\beae{\label{eq. vector chiral cross correlation}
F_{\cV V}  (\muk, \eta, \eta^\prime, k) = ~& -\frac{i}{4} k^2 \muk \left(1-\muk^4\right)
\left\{c_{+2}^2 P^{(+2)} (\eta, \eta^\prime, k) - c_{-2}^2 P^{(-2)} (\eta, \eta^\prime, k)\right\}
\\
&
- \frac{i}{2} k^2 \muk \left(1 - \muk^2\right)^2 
\left\{c_{+1}^2 P^{(+1)} (\eta, \eta^\prime, k) - c_{-1}^2 P^{(-1)} (\eta, \eta^\prime, k)\right\}
~.
}
\subsection{Projected scalar field}
Let us move on to scalar quantities. Due to the presence of the trace of the projected tensor field, one can introduce three different types of scalar fields as
\bae{
\psi (\eta, \bm{k}) \coloneqq g^{AB} \hat{h}_{AB} (\eta, \bm{k}) ~, \quad 
S (\eta, \bm{k}) \coloneqq ik^A V_A (\eta, \bm{k}) ~, \quad 
\cS (\eta, \bm{k}) \coloneqq ik^A \cV_A (\eta, \bm{k}) ~.
}
The first one is a trace component of the projected tensor field, whose power spectrum is defined by
\bae{
\Braket{\psi (\eta, \bm{k}) \psi (\eta^\prime, \bm{k}^\prime)} = \left(2\pi\right)^3\DD  (\bm{k}+\bm{k}^\prime) F_{\psi\psi} (\muk, \eta, \eta^\prime, k) ~,
}
with
\beae{
F_{\psi\psi} (\muk, \eta, \eta^\prime, k) = ~&
\frac14 \left(1-\muk^2\right)^2 
\left\{c_{+2}^2 P^{(+2)} (\eta, \eta^\prime, k) + c_{-2}^2 P^{(-2)} (\eta, \eta^\prime, k)\right\}
\\
&
+ \muk^2 \left(1 - \muk^2\right) 
\left\{c_{+1}^2 P^{(+1)} (\eta, \eta^\prime, k) + c_{-1}^2 P^{(-1)} (\eta, \eta^\prime, k)\right\}
\\
& + \frac12 \left(1 + \muk^2\right)^2 c_{b}^2 P^{(b)}(\eta, \eta^\prime, k)
+ \left(1 - \muk^2\right)^2 c_{\ell}^2 P^{(\ell)}(\eta, \eta^\prime, k)
~.
}
The other auto power spectra are defined respectively as
\bae{
\Braket{S (\eta, \bm{k}) S (\eta^\prime, \bm{k}^\prime)} = \left(2\pi\right)^3\DD  (\bm{k}+\bm{k}^\prime) F_{SS} (\muk, \eta, \eta^\prime, k) ~,
}
with
\beae{\label{eq. scalar auto power}
F_{SS} (\muk, \eta, \eta^\prime, k) = ~&
\frac14 k^4 \muk^4 \left(1-\muk^2\right)^2
\left\{c_{+2}^2 P^{(+2)} (\eta, \eta^\prime, k) + c_{-2}^2 P^{(-2)} (\eta, \eta^\prime, k)\right\}
\\
& 
+ k^4 \muk^2 \left(1 - \muk^2\right)^3 
\left\{c_{+1}^2 P^{(+1)} (\eta, \eta^\prime, k) + c_{-1}^2 P^{(-1)} (\eta, \eta^\prime, k)\right\}
\\
& + \frac12 k^4 \muk^4\left(1 - \muk^2\right)^2 c_{b}^2 P^{(b)}(\eta, \eta^\prime, k)
+ k^4 \left(1 - \muk^2\right)^4 c_{\ell}^2 P^{(\ell)}(\eta, \eta^\prime, k)
~,
}
and 
\bae{
\Braket{\cS (\eta, \bm{k}) \cS (\eta^\prime, \bm{k}^\prime)} = \left(2\pi\right)^3\DD  (\bm{k}+\bm{k}^\prime) F_{\cS\cS} (\muk, \eta, \eta^\prime, k) ~,
}
with
\beae{\label{eq. curl scalar auto power}
F_{\cS\cS} (\muk, \eta, \eta^\prime, k) = ~&
\frac14 k^4 \muk^2 \left(1-\muk^2\right)^2
\left\{c_{+2}^2 P^{(+2)} (\eta, \eta^\prime, k) + c_{-2}^2 P^{(-2)} (\eta, \eta^\prime, k)\right\}
\\
&
+ \frac14 k^4\left(1 - \muk^2\right)^3 
\left\{c_{+1}^2 P^{(+1)} (\eta, \eta^\prime, k) + c_{-1}^2 P^{(-1)} (\eta, \eta^\prime, k)\right\}
~.
}

There can be three types of cross-correlations. The cross-correlations with the trace are given by
\bae{
\Braket{\psi (\eta, \bm{k}) S (\eta^\prime, \bm{k}^\prime)} & = \left(2\pi\right)^3\DD  (\bm{k}+\bm{k}^\prime) F_{\psi S} (\muk, \eta, \eta^\prime, k) ~,
\\
\Braket{\psi (\eta, \bm{k}) \cS  (\eta^\prime, \bm{k}^\prime)} & = \left(2\pi\right)^3\DD  (\bm{k}+\bm{k}^\prime) F_{\psi\cS} (\muk, \eta, \eta^\prime, k) ~,
}
where their kernel functions are
\beae{\label{eq. scalar cross correlation 1}
F_{\psi S} (\muk, \eta, \eta^\prime, k) 
= ~& 
 \frac{1}{4} k^2 \muk^2 \left(1-\muk^2\right)^2 
 \left\{c_{+2}^2 P^{(+2)} (\eta, \eta^\prime, k) + c_{-2}^2 P^{(-2)} (\eta, \eta^\prime, k)\right\}
 \\
 &
- \frac{1}{2} k^2 \muk^2 \left(1 -\muk^2\right)^2 
\left\{c_{+1}^2 P^{(+1)} (\eta, \eta^\prime, k) + c_{-1}^2 P^{(-1)} (\eta, \eta^\prime, k)\right\}
\\
& - \frac12 k^2\muk^2\left(1 - \muk^4\right) c_{b}^2 P^{(b)}(\eta, \eta^\prime, k)
- k^2 \left(1 - \muk^2\right)^3 c_{\ell}^2 P^{(\ell)}(\eta, \eta^\prime, k)
~,
}
and
\beae{\label{eq. scalar cross correlation 2}
F_{\psi \cS} (\muk, \eta, \eta^\prime, k) 
= ~& 
\frac{i}{4} k^2 \muk \left(1-\muk^2\right)^2 
\left\{c_{+2}^2 P^{(+2)} (\eta, \eta^\prime, k) - c_{-2}^2 P^{(-2)} (\eta, \eta^\prime, k)\right\}
\\
&
- \frac{i}{2} k^2 \muk \left(1 - \muk^2\right)^2 
\left\{c_{+1}^2 P^{(+1)} (\eta, \eta^\prime, k) - c_{-1}^2 P^{(-1)} (\eta, \eta^\prime, k)\right\}
~.
}
The remaining cross-correlation is
\bae{
\Braket{\cS (\eta, \bm{k}) S (\eta^\prime, \bm{k}^\prime)} & = \left(2\pi\right)^3\DD  (\bm{k}+\bm{k}^\prime) F_{\cS S} (\muk, \eta, \eta^\prime, k) ~,
}
with
\beae{\label{eq. scalar cross correlation 3}
F_{\cS S} (\muk, \eta, \eta^\prime, k) 
= ~& 
- \frac{i}{4} k^4 \muk^3 \left(1-\muk^2\right)^2 
\left\{c_{+2}^2 P^{(+2)} (\eta, \eta^\prime, k) - c_{-2}^2 P^{(-2)} (\eta, \eta^\prime, k)\right\}
\\
&
- \frac{i}{2} k^4 \muk \left(1 - \muk^2\right)^3 
\left\{c_{+1}^2 P^{(+1)} (\eta, \eta^\prime, k) - c_{-1}^2 P^{(-1)} (\eta, \eta^\prime, k)\right\}
~.
}
\section{Relation to the $E/B$-mode decomposition}\label{sec. EB correlation}
Let us show how our formalism is related to that of the $E/B$-modes shown in refs.~\cite{Akitsu:2022lkl,Philcox:2023uor,Saga:2023afb,Okumura:2024xnd}. 
Under the flat-sky approximation, the  $E$-mode and $B$-mode are defined by
\bae{
  E (\bm{k}, \hat{n}) \pm i B (\bm{k}, \hat{n}) \coloneqq m^i_{\mp}(\hat{n}) m^j_{\mp}(\hat{n}) t^{\mathrm{T}}_{ij} (\bm{k}) \ee^{\mp 2 i \phi_k} ~,
}
where $t^{\mathrm{T}}_{ij}$ is the traceless galaxy shape field and $m^i_{\mp}(\hat{n})$ are the basis vectors defined by
\bae{
  \bm{m}_\pm (\hat{n}) = \frac{1}{\sqrt{2}} \left(1, \mp 1, 0 \right) ~.
}
The three-dimensional power spectra, observables in galaxy redshift surveys, are defined by
\bae{
  \Braket{X (\eta, \bm{k}) Y (\eta^\prime, \bm{k}^\prime)} = \left(2\pi\right)^3\DDD (\bm{k}+\bm{k}^\prime) P_{XY} (\muk, \eta, \eta^\prime, k) ~, \quad X,Y \in \{E,B\} ~.
}
Recall that the dimensionless traceless galaxy shape field is given by eq.~\eqref{eq. shape field and tensor} and it can be expressed as
\bae{
  t^{\mathrm{T}}_{ij} = - C (\eta, k) \hhT_{ij} ~.
}
Then the $E$-mode and $B$-mode power spectra are given by
\bae{
  &\begin{aligned}
    P_{EE} / C(\eta, k) C(\eta^\prime, k) = ~& \frac{1}{16}\left(1+\muk^2\right)^2 \left\{c_{+2}^2 P^{(+2)} (\eta, \eta^\prime, k) + c_{-2}^2 P^{(-2)} (\eta, \eta^\prime, k)\right\}
    \\
    & + \frac{1}{4}\muk^2\left(1-\muk^2\right) \left\{c_{+1}^2 P^{(+1)} (\eta, \eta^\prime, k) + c_{-1}^2 P^{(-1)} (\eta, \eta^\prime, k)\right\}
    \\
    & + \frac{1}{8}\left(1-\muk^2\right)^2 c_{b}^2 P^{(b)}(\eta, \eta^\prime, k) + \frac{1}{4}\left(1-\muk^2\right)^2 c_{\ell}^2 P^{(\ell)}(\eta, \eta^\prime, k)
  ~,
  \end{aligned}
  \\
  &
  \begin{aligned}
  P_{BB} / C(\eta, k) C(\eta^\prime, k) = ~& \frac{1}{4}\muk^2 \left\{c_{+2}^2 P^{(+2)} (\eta, \eta^\prime, k) + c_{-2}^2 P^{(-2)} (\eta, \eta^\prime, k)\right\}
  \\
  & + \frac{1}{4}\left(1-\muk^2\right)\left\{c_{+1}^2 P^{(+1)} (\eta, \eta^\prime, k) + c_{-1}^2 P^{(-1)} (\eta, \eta^\prime, k)\right\} ~,
  \end{aligned}
}
while the cross-power spectra takes the form
\baea{
  P_{EB} / C(\eta, k) C(\eta^\prime, k) = ~&
  \frac{i}{8}\muk \left(1+\muk^2\right)\left\{c_{+2}^2 P^{(+2)} (\eta, \eta^\prime, k) - c_{-2}^2 P^{(-2)} (\eta, \eta^\prime, k)\right\}
  \\
  & + \frac{i}{4}\muk\left(1-\muk^2\right)\left\{c_{+1}^2 P^{(+1)} (\eta, \eta^\prime, k) - c_{-1}^2 P^{(-1)} (\eta, \eta^\prime, k)\right\} ~.
}
By comparing the expressions derived in section~\ref{sec. Projected vector perturbation} and \ref{sec. Projected scalar perturbation}, one sees that these $EE$-, $BB$-, and $EB$-spectra carry the same information as spectra consisting of the projected tensor fields, namely,
\bae{
  F^{\mathrm{T}}_{S S} (\muk, \eta, \eta^\prime, k) 
  & = k^4 \left(1-\muk^2\right)^2 P_{EE} /  C(\eta, k) C(\eta^\prime, k) ~,
  \\
  F^{\mathrm{T}}_{\cV V} (\muk, \eta, \eta^\prime, k) & =  - 2 k^2 \left(1-\muk^2\right) P_{EB} / C(\eta, k) C(\eta^\prime, k)  ~,
  \\
  F^{\mathrm{T}}_{\cS\cS} (\muk, \eta, \eta^\prime, k) & = k^4 \left(1-\muk^2\right)^2 P_{BB} / C(\eta, k) C(\eta^\prime, k) ~.
}

Let us briefly comment on the detectability using a bias function considered in refs.~\cite{Schmidt:2013gwa,Akitsu:2022lkl} and focusing on equal-time correlations. We assume that the tidal field is given by
\bae{\label{eq. shape bias EB}
  t_{ij} = b_{\mathrm{K}}^{\mathrm{GW}}(\eta, k) \hat{h}_{ij} (\eta_\ini, \bm{k}) ~,
}
and the linear shape bias $b_{\mathrm{K}}^{\mathrm{GW}} (\eta, k)$ is given by
\bae{
  b_{\mathrm{K}}^{\mathrm{GW}} (\eta,k) =\frac74 \alpha (\eta,k) b_{\mathrm{K}}^{\mathrm{Scalar}} ~,
}
where $b_{\mathrm{K}}^{\mathrm{Scalar}}$ is the linear shape bias induced by the scalar tides and $\alpha (\eta,k)$ exhibits the time evolution of a mode. For modes that enter the horizon during the matter-dominated era, one can obtain an analytical form of $\alpha (\eta,k)$ as
\bae{
  \alpha_{\mathrm{MD}} (\eta, k) \coloneqq \frac45 \left[\beta (\eta, k) - \int_0^\eta \dd\tilde{\eta} \left(\frac{\tilde{\eta}}{\eta}\right)^5 \frac{\dd \beta (\tilde{\eta}, k)}{\dd \tilde{\eta}}\right] ~, \quad 
  \beta (\eta, k) \coloneqq \frac12 \left[1-3\frac{j_1 (k\eta)}{k\eta}\right] ~,
}
with $j_i$ being the spherical Bessel function.\footnote{Ref.~\cite{Akitsu:2022lkl} focuses on \ac{GWs} generated during inflation and equal-time correlations.} For modes with higher $k$, it is difficult to obtain an analytical expression and one needs to rely on numerical calculation (see ref.~\cite{Akitsu:2022lkl} for details). However, according to ref.~\cite{Akitsu:2022lkl}, the scale dependence of the shape bias for higher $k$ is not significant in simulations. Thus we simply assume
\bae{
  \alpha(\eta, k) = \alpha_{\mathrm{MD}} (\eta, k) \Theta (k - 10^{-3} ~ h/\mathrm{Mpc}) +  \alpha_{\mathrm{MD}} (\eta, 10^{-3} ~ h/\mathrm{Mpc}) \Theta (10^{-3} ~ h/\mathrm{Mpc} - k) ~,
}
with $\Theta$ being the step function. We note that the shape bias introduced in eq.~\eqref{eq. shape bias EB} relates the tidal field with the initial tensor perturbation. Hence, in our expressions, we replace
\bae{
    C^2 (\eta, k) P^{(\lambda)}(\eta, \eta^\prime, k) \to \left(b_{\mathrm{K}}^{\mathrm{GW}} (\eta, k)\right)^2 P^{(\lambda)}_{\mathrm{ini}} (k) ~,
}
for equal-time correlations.

Let us expand the $BB$- and $EB$-spectra in terms of the Legendre polynomials $\calP_\ell (\mu_k)$ as
\bae{
  P_{X, \ell} (k) = \frac12 \int^{1}_{-1}\dd\mu_k \calP_\ell (\mu_k) P_{X} (\mu_k, k) ~,  \quad X \in \{BB, EB\} ~.
}
The monopole and quadrupole of the $BB$-power spectrum are given by
\bae{
P_{BB, 0}/\left(b_{\mathrm{K}}^{\mathrm{GW}} (\eta, k)\right)^2 & = 
  \frac{1}{12} \left(c_{+2}^2 P^{(+2)}_{\mathrm{ini}} + c_{-2}^2 P^{(-2)}_{\mathrm{ini}}\right)
+ \frac{1}{6} \left(c_{+1}^2 P^{(+1)}_{\mathrm{ini}} + c_{-1}^2 P^{(-1)}_{\mathrm{ini}}\right) ~,
\\
P_{BB, 2} /\left(b_{\mathrm{K}}^{\mathrm{GW}} (\eta, k)\right)^2 & = 
  \frac{1}{30} \left(c_{+2}^2 P^{(+2)}_{\mathrm{ini}} + c_{-2}^2 P^{(-2)}_{\mathrm{ini}}\right)
- \frac{1}{30} \left(c_{+1}^2 P^{(+1)}_{\mathrm{ini}} + c_{-1}^2 P^{(-1)}_{\mathrm{ini}}\right) ~.
}
For the $EB$ cross-power spectrum, its dipole and octupole take the form
\bae{
P_{EB, 1} /\left(b_{\mathrm{K}}^{\mathrm{GW}} (\eta, k)\right)^2 & =
  \frac{i}{15} \left(c_{+2}^2 P^{(+2)}_{\mathrm{ini}} - c_{-2}^2 P^{(-2)}_{\mathrm{ini}}\right)
+ \frac{i}{30} \left(c_{+1}^2 P^{(+1)}_{\mathrm{ini}} - c_{-1}^2 P^{(-1)}_{\mathrm{ini}}\right) ~,
\\
P_{EB, 3} /\left(b_{\mathrm{K}}^{\mathrm{GW}} (\eta, k)\right)^2 & =
  \frac{i}{140} \left(c_{+2}^2 P^{(+2)}_{\mathrm{ini}} - c_{-2}^2 P^{(-2)}_{\mathrm{ini}}\right)
- \frac{i}{70} \left(c_{+1}^2 P^{(+1)}_{\mathrm{ini}} - c_{-1}^2 P^{(-1)}_{\mathrm{ini}}\right) ~.
}

\begin{figure}[tb] 
  \centering
  \begin{minipage}[b]{0.49\linewidth}
    \centering
    \includegraphics[width=\linewidth]{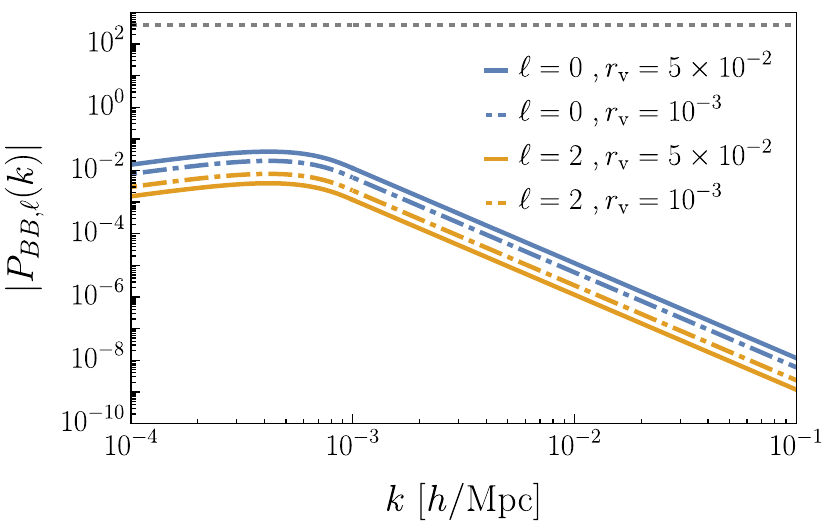}
  \end{minipage}
  \hfill
  \begin{minipage}[b]{0.49\linewidth}
    \centering
    \includegraphics[width=\linewidth]{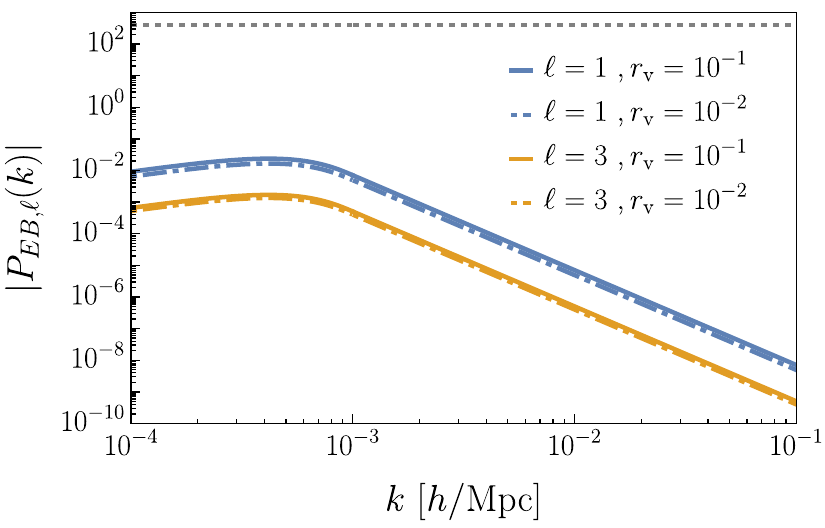}
  \end{minipage}
  \caption{The first two multipoles of the $BB$ (left) and $EB$ (right) power spectrum at $z=1$ assuming $r=0.1$, $b_{\mathrm{K}}^{\mathrm{Scalar}}=0.1$, and $c_\lambda=1$. For the $EB$ cross power spectrum, we assume the maximal chirality. The gray dashed line shows the shape noise $\sigma^2_\gamma/\bar{n}_g$, where we take $\sigma^2_\gamma =0.2$ and $\bar{n}_g=5 \times 10^{-4} ~ (h/\mathrm{Mpc})^3$.
}\label{fig: EB power}
\end{figure}

As an example, we consider scale-free spectra for the initial power spectra:
\bae{
  P^{(\mathrm{T})}_{\mathrm{ini}} = \frac{2\pi^2}{k^3} r A_s ~, \quad 
  P^{(\mathrm{V})}_{\mathrm{ini}} = \frac{2\pi^2}{k^3} r_{\mathrm{v}} A_s ~,
}
where $A_s = 2.1 \times 10^{-9}$, $r$ is the tensor-to-scalar ratio, and $r_{\mathrm{v}}$ is the vector-to-scalar ratio. 
In figure~\ref{fig: EB power}, we plot the multipole moments of the $BB$- and $EB$- spectra assuming $r=0.1$, $b_{\mathrm{K}}^{\mathrm{Scalar}}=0.1$, and $c_\lambda=1$ at $z=1$. We also show the Poisson shape noise $\sigma^2_\gamma/\bar{n}_g$ with $\sigma^2_\gamma =0.2$ and $\bar{n}_g=5 \times 10^{-4} ~ (h/\mathrm{Mpc})^3$. The limit $r_{\mathrm{v}}\to0$ corresponds to figure~12 of ref~\cite{Akitsu:2022lkl}.
Given the amplitude of the \ac{GWs} and the shape noise, one may think that it is challenging to detect GWs using the galaxy-shape measurements. However, at the linear order, there is a possibility of generating \ac{GWs} with a higher amplitude than those originating from inflation at different scales~\cite{Gorji:2023ziy}. Furthermore, optimizing the \ac{IAs} signal through luminosity weighting to brighter galaxies (e.g., ref.~\cite{Seljak:2009af}) may further enhance sensitivity.
\section{Different way of calculating the angular integral}\label{sec. the angular integral}
While we used the spherical harmonics to perform the angular integral of the \ac{ORF} in section~\ref{sec. ORF}, it can be analyzed directly as done in ref.~\cite{Okumura:2024xnd}. Here, with one example, we show that these two ways of calculation give the same result.

\begin{figure}[tb]
  \centering
  \includegraphics[width=.8\hsize]{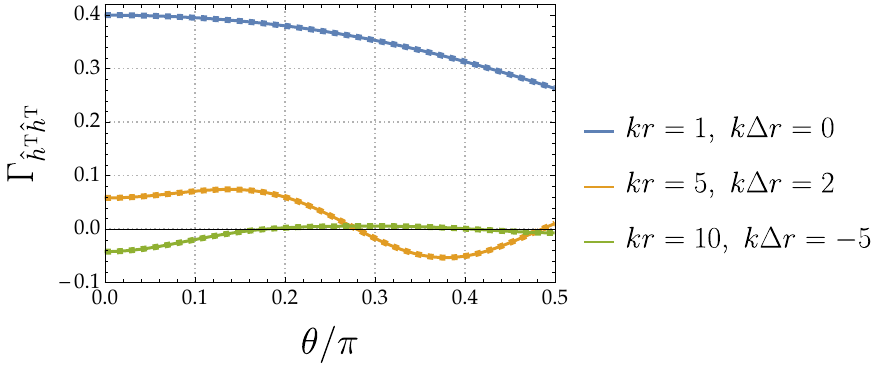}
  \caption{The overlap reduction function obtained from the direct integration of eq.~\eqref{eq. direct htht} (solid line) and the spherical harmonics expansion \eqref{eq. ORF YLM} with $s=v=0$ (dashed line). With several parameters, we check that the two ways give the same results.
  }
  \label{Fig. PlotTwoWays}
\end{figure}

Let us consider the following two-point function
\bae{
\Gamma_{\hhT\hhT} = \cos\left(k\Delta r\right) \int\dd\Omega_k \frac18\left(1+6 \muk^2 + \muk^4 \right) \ee^{i\bm{k}\cdot \left(\bm{x}-\bm{x}^\prime\right)} ~.
}
With the coordinate system
\bae{
  x^\prime{}^\mu = \left( \eta^\prime, 0, 0, r^\prime \right) ~, 
  \quad 
  x{}^\mu = \left( \eta, r \sin\theta \cos\phi, r\sin\theta \sin\phi, r\cos\theta \right) ~,
}
and the wavevector
\bae{
  \bm{k} = \left( k \sin\theta_k \cos\phi_k, k \sin\theta_k \sin\phi_k, k\cos\theta_k \right) ~,
}
the exponential function reduces to
\bae{
\ee^{i\bm{k}\cdot \left(\bm{x}-\bm{x^\prime}\right)} = \ee^{ikr\muk \left\{\cos{\theta} -\left(1+\frac{k\Delta r}{kr}\right)\right\}}\ee^{ikr\cos(\phi-\phi_k)\sin{\theta}\sin{\theta_k}} ~.
}
The $\phi_k$-integral can be performed analytically as
\bae{
\int_{0}^{2\pi} \dd \phi_k \ee^{i\bm{k}\cdot \left(\bm{x}-\bm{x^\prime}\right)} = 2\pi \exp[ikr\muk \left\{\cos{\theta} -\left(1+\frac{k\Delta r}{kr}\right)\right\}] J_0 \left(kr \sqrt{1-\muk^2} \sin{\theta} \right) ~,
}
where $J_0$ is the zeroth order Bessel function of first kind. We eventually find that the \ac{ORF} takes the form
\bae{\label{eq. direct htht}
\Gamma_{\hhT\hhT} = \cos\left(k\Delta r\right) \int_1^1 \dd\muk \frac{1+6\muk^2+\muk^4}{16}\ee^{-i\muk \{k\Delta r + 2kr\sin^2\left(\theta/2\right)\}} J_0 \left(kr \sqrt{1-\muk^2} \sin{\theta} \right) ~.
}
Figure~\ref{Fig. PlotTwoWays} shows the above \ac{ORF} with different parameters, implying that the two ways of calculations agree each other.  

\newpage

\def\arxivfont{\rm}
\bibliographystyle{JCAP}
\baselineskip=.95\baselineskip
\let\originalthebibliography\thebibliography
\renewcommand\thebibliography[1]{
\originalthebibliography{#1}
\setlength{\itemsep}{0pt plus 0.3ex}
}
\bibliography{Bib}

\end{document}